\documentclass[journal]{IEEEtran}

\usepackage[algo2e]{algorithm2e} 
\usepackage{amsmath,amssymb,amsmath,amsfonts}
\usepackage{bm}
\usepackage{bbm}
\usepackage{graphicx}
\usepackage{cite}
\usepackage{epstopdf}
\usepackage{gensymb}
\usepackage{color}
\usepackage{lipsum}
\usepackage{float}
\usepackage{epstopdf}
\usepackage[mathcal]{eucal}
\usepackage{subfiles}
\usepackage[T1]{fontenc}
\usepackage{siunitx}
\usepackage{tabulary}
\usepackage{epsfig,graphics,graphicx,subfig,amssymb,amstext,amsmath,algorithm,algorithmic,wrapfig,multirow}
\usepackage{array}
\usepackage{adjustbox}
\usepackage[dvipsnames]{xcolor}
\usepackage{cite}
\usepackage{times}
\usepackage{placeins}
\usepackage{textcomp}
\usepackage[font=small]{caption}
\usepackage{flushend}
\usepackage{color, colortbl}
\usepackage{tabularx}
\usepackage{bm}
\usepackage{enumerate}
\usepackage{tikz}
\usepackage{pgfplots}
\usepackage{epstopdf}
\usetikzlibrary{shapes,arrows}
\usepackage[utf8]{inputenc}
\usepackage{mathtools}
\usepackage{xcolor}
\usepackage{tikz}
\usetikzlibrary{chains,arrows,calc,positioning}
\usepackage{amssymb}
\usepackage{pifont}
\usepackage{soul}
\usepackage{breqn} 
\usepackage{booktabs} 
\usepackage{multirow}
\usepackage{enumitem}
\usepackage{tabulary}
\usepackage[normalem]{ulem}
\usepackage{dblfloatfix}
\usepackage{makecell}
\usepackage{lipsum}
\usepackage{amsthm}
\usepackage{comment}
\usepackage{subfig}
\usepackage{filecontents}
\usepackage{url}
\usepackage{bbm}

\newcommand{\angel}[1]{\noindent{\textsf{[Angel]: {\color{red}#1}}} }

\newtheoremstyle{mystyle}
  {}
  {}
  {\itshape}
  {}
  {\bfseries}
  {.}
  { }
  {}
\theoremstyle{mystyle}

\newlength \figwidth
\definecolor{bittersweet}{rgb}{1.0, 0.44, 0.37}
\definecolor{glaucous}{rgb}{0.38, 0.51, 0.71}
\definecolor{gainsboro}{rgb}{0.86, 0.86, 0.86}
\definecolor{babyblueeyes}{rgb}{0.63, 0.79, 0.95}
\definecolor{silver}{rgb}{0.75, 0.75, 0.75}
\definecolor{neoncarrot}{rgb}{1.0, 0.64, 0.26}
\definecolor{Gray}{gray}{0.9}
\definecolor{LightCyan}{rgb}{0.88,1,1}
\definecolor{BackgroundLightBlue}{rgb}{0.97,0.97,1}
\definecolor{BackgroundGray}{gray}{0.98}

\newcommand{\blue}[1]{\textcolor{blue}{#1}}
\newcommand{\red}[1]{{\textcolor[rgb]{1,0,0}{#1}}}

\makeatletter

 \let\oldforeign@language\foreign@language
 \DeclareRobustCommand{\foreign@language}[1]{%
   \lowercase{\oldforeign@language{#1}}}

\def\nb0{{\mathbf{0}}}
\def\nb1{{\mathbf{1}}}

\makeatother

\IEEEoverridecommandlockouts
\begin{document}

\bstctlcite{IEEEexample:BSTcontrol}

\title{Data-Driven Deployment of Reconfigurable Intelligent Surfaces in Cellular Networks}

\author{\IEEEauthorblockN{{Sina Beyraghi, \emph{Graduate Student Member, IEEE}, Javad Shabanpour, \emph{Graduate Student Member, IEEE},\\ Giovanni Geraci, \emph{Senior Member, IEEE}, Paul Almasan, and Angel Lozano, \emph{Fellow, IEEE}} 
\thanks{S. Beyraghi is with Telefónica Research and Universitat Pompeu Fabra, Spain. \emph{(email: mohammadsina.beyraghi@telefonica.com)}}
\thanks{J. Shabanpour is with Nokia Bell Labs and Aalto University, Finland. He was with Telefónica Research, Spain, when this work was carried out.}
} 
\thanks{G. Geraci is with Nokia Standards and Universitat Pompeu Fabra, Spain. He was with Telefónica Research, Spain, when this work was carried out.}
\thanks{P. Almasan is with Telefónica Research, Spain.}
\thanks{A. Lozano is with Universitat Pompeu Fabra, Spain.}
\thanks{This work was in part supported by H2020-MSCA-ITN-2020 META WIRELESS (Grant Agreement: 956256), by the SNS
JU Horizon Europe Project under Grant Agreement No.
101139161 (INSTINCT), by the 6G-Machine Intelligence based Radio Access Infrastructure (6G-MIRAI) project under Grant Agreement No. 101192369, by the Spanish Research Agency through grants PID2021-123999OB-I00 and CNS2023-145384, by ICREA, by the Maria de Maeztu Units of Excellence Programme (CEX2021-001195-M), and by the Spanish Ministry of Economic Affairs and Digital Transformation and the European Union NextGenerationEU through UNICO-5G I+D projects TSI-063000-2021-138 (SORUS-RIS) and TSI-063000-2021-59 (RISC-6G).}
\thanks{Some of these results will be presented at IEEE Globecom 2025~\cite{BeyShaGer2025sina}.}
}

\maketitle

\begin{abstract}

This paper presents a fully automated, data-driven framework for the large-scale deployment of reconfigurable intelligent surfaces (RISs) in cellular networks. Leveraging physically consistent ray tracing and empirical data from a commercial deployment in the UK, the proposed method jointly optimizes RIS placement, orientation, configuration, and base station beamforming in dense urban environments across frequency bands (corresponding to 4G, 5G, and a hypothetical 6G system). Candidate RIS locations are identified via reflection- and scattering-based heuristics using calibrated electromagnetic models within the Sionna Ray Tracing (RT) engine. Outage users are clustered to reduce deployment complexity, and the tradeoff between coverage gains and infrastructure cost is systematically evaluated. It is shown that achieving meaningful coverage improvement in urban areas requires a dense deployment of large-aperture RIS units, raising questions about cost-effectiveness. To facilitate reproducibility and future research, the complete simulation framework and RIS deployment algorithms are provided as open-source software.\footnote{Available at: \url{https://github.com/Telefonica-Scientific-Research/DDRD}} 

\end{abstract}


\section{Introduction}
\label{sec:Sec1}

For mobile network operators (MNOs), consistent connectivity, reduced deployment costs, and improved energy efficiency are critical concerns, especially in dense urban environments where obstructions, building materials, and multipath propagation can severely degrade coverage~\cite{OugGerPol2025}. 
These challenges become even more pronounced at higher frequencies, where increased pathloss and weaker diffraction lead to signal dead zones and reduced link reliability. Reconfigurable intelligent surfaces (RIS) have emerged as a promising technology to address some of these limitations in future-generation networks~\cite{9140329}. By intelligently reflecting signals toward underserved areas without requiring additional transmit power, RIS offer a potentially energy-efficient means of enhancing coverage~\cite{8741198}. However, it remains unclear whether MNOs will adopt RIS at scale, given the uncertainty surrounding their cost-vs-performance tradeoffs.

\subsection{Motivation and Related Work}

The integration of RIS in wireless networks presents major challenges, particularly in dense urban settings. 
The problem of RIS placement is inherently high-dimensional and combinatorial, involving an exploding number of potential sites, orientation constraints, and intricate environment-specific propagation effects. Trial-and-error approaches have poor performance and are typically infeasible at scale.
In addition to placement, determining parameters such as the quantity of RIS units and their aperture size is equally crucial. Arbitrary configurations may fall short performance-wise or lead to unnecessary capital and operational expenditures for MNOs. These complexities highlight the need for an automated and scalable framework for RIS deployment and performance evaluation in realistic urban scenarios.


A number of studies have explored RIS placement and configuration strategies, primarily based on analytically tractable models and simplified geometries. 
In~\cite{9201413}, the authors optimize the direction and horizontal distance of the RIS to improve coverage, in a setting with one BS and one user equipment (UE). Similarly,~\cite{9326394} investigates the optimal placement of a single RIS in a simplified setup involving a single-antenna transmitter and receiver; the ensuing analysis indicates that the RIS should be positioned next to either the transmitter or the receiver. While theoretically insightful, such findings have limited applicability in dense urban settings with multiple UEs and complex building layouts. 

More elaborate studies 
such as~\cite{9530750, 9712623} optimize the RIS phase shifts and placement to minimize path loss in single-antenna communication. For the case of multiple transmit antennas,~\cite{9351782, 9852464} propose a 3D beam broadening and flattening approach based solely on line-of-sight (LoS) components to guide RIS positioning. In~\cite{10211255}, RIS placement and orientation for directional multicasting are modeled as a multilevel facility location problem, applying a branch-and-bound algorithm. 
Several studies~\cite{9745477, 9884994, 9586067} further confirm that proper RIS placement can improve transmission rates. However, all of these works still consider a single BS and a single RIS, rather than multicell multi-RIS networks.

One study~\cite{9852389} does apply
system-level simulations to demonstrate that RIS units can enhance coverage: from 77\% to 95\% at sub-6\,GHz frequencies (FR1), and from 46\% to 95\% at mmWave frequencies (FR2). Although this work considers a multicell multi-UE multi-RIS scenario, the RIS are positioned randomly without any optimization strategy. 
A more recent study~\cite{10558715} proposes a rule-based and branch-and-bound method for RIS location optimization, aiming at balancing coverage and cost
on the basis of stochastic channel models that do not account for site-specific features. In \cite{10.1145/3629134}, several tall buildings are selected and RIS units are installed on their facades to cover outage UEs, but a method to deploy RISs in a large-scale network is not proposed. For indoor scenarios, \cite{9838688} focuses on deploying and configuring RISs using ray tracing-based propagation modeling, yet identifying suitable locations for RISs is generally less challenging indoors. 

\subsubsection{RIS Density}

Previous work has shown that the required density of RIS units depends strongly on the density of environmental blockages~\cite{9174910}. For example, achieving a blind-spot outage probability below $10^{-5}$ may require on the order of six RIS units per km$^2$ in areas with 300 blockages/km$^2$, but up to 490 RIS units per km$^2$ in environments with 700 blockages/km$^2$. These estimates are based on stochastic channel models that ignore site-specific features. 
As an alternative, ray tracing evaluations have been proposed in~\cite{10279515, electronics12051173}, but these studies are limited to a single BS indoors. 

\subsubsection{Number of Elements and Apertures}

Another important factor that directly affects the RIS performance is the number of reflecting elements per RIS unit. This determines the RIS aperture given an element spacing. While increasing the number of elements is bound to enhance signal reflection and reduce the required transmit power, it also results in higher hardware complexity, increased channel estimation overhead, and greater fabrication costs. Consequently, the number of RIS elements must be carefully optimized~\cite{8888223, 9352958, 9852735}.

\subsubsection{RIS Reflection Model}

A widely adopted approach to model RIS behavior is the diagonal phase-shift matrix
\begin{equation}
\mathbf{\Theta} = \mathrm{diag} \! \left(  A_1 e^{j\varphi_1},  A_2 e^{j\varphi_2}, \ldots,  A_N e^{j\varphi_N} \right),
\label{eq:RIS_phase_matrix}
\end{equation}
where\( \ A_n \) denotes the magnitude reflection factor and \( \varphi_n \) the phase shift introduced by the \( n \)th RIS element~\cite{8811733, 8796365, 9140329}. While analytically convenient, this model posits an idealized behavior that neglects important physical effects, such as angle-dependent scattering or element interactions. Consequently, it fails to capture performance degradation at large deflection angles.
Even with improved parameterizations, \eqref{eq:RIS_phase_matrix} cannot accurately predict the reduction in reflection efficiency at high angular deviations~\cite{10473672}. These inaccuracies are problematic in urban settings, where non-LoS (NLoS) paths with high deflection angles are common. 
Recent efforts have proposed ray tracing to integrate RIS into system simulations~\cite{sandh2024ray, 10301521}, yet most such ray tracers lack electromagnetic consistency: they do not account for polarization effects or energy conservation principles. To address this shortcomings, the present work adopts a full-wave-inspired ray-tracing-compatible model that characterizes the RIS as a programmable Huygens surface. Building on the macroscopic formulation in~\cite{9713744, 10419169}, an efficient ray-based implementation is leveraged that enables decomposition of the reradiated field into specular, anomalous, and diffuse components. 
This model is integrated into a 3D simulator via the open-source  Sionna Ray Tracing (RT) framework (v0.19.2) framework~\cite{hoydis2023sionna}, which supports, at an affordable computational cost, scalable ray tracing simulations that are physically grounded.

\subsubsection{Summary of Gaps}
While existing studies have provided valuable insights into RIS placement, reflection modeling, and aperture design, they suffer from several key limitations. Most works are confined to simplified single-UE or single-cell scenarios, relying on idealized or stochastic channel models that overlook site-specific attributes~\cite{9530750, 9712623, 9351782, 9852464, 10211255}. Others adopt oversimplified reflection models that fail to capture polarization and angular-dependent efficiency loss at high deflection angles~\cite{8811733, 8796365, 9140329, 10473672, sandh2024ray, 10301521, BeyShaGer2025}. Physical constraints (e.g., building geometry, materials, and feasible orientations) are often ignored~\cite{9745477, 9884994, 9586067}. Only a few studies integrate ray tracing with empirical data, and none offer a scalable, automated solution across different frequencies and network topologies. These gaps underscore the need for a framework to support end-to-end deployment and evaluation of RIS-assisted networks in complex urban environments.

\subsection{Approach and Contribution}

This paper sets forth a fully automated data-driven framework for the large-scale deployment of RIS in cellular networks. This framework can assist in determining key parameters (e.g., RIS placement, orientation, phase-shift configuration, and BS beamforming) based on site-specific propagation modeling. In contrast to trial-and-error methods and state-of-the-art analytical approaches, the proposed framework is designed for large-scale multicell networks and multiple frequency bands and cellular generations. Crucially, it integrates a calibrated ray-tracing engine to accurately capture electromagnetic interactions with specific materials and geometries.

The data-driven framework is 
tested with data from a real-world production deployment from a leading commercial cellular operator in UK. Simulations are conducted for 4G, 5G, and a hypothetical 6G system, respectively operating at $2$\,GHz, $3.5$\,GHz, and $10$\,GHz, using the open-source Sionna RT engine~\cite{hoydis2023sionna}. Materials are calibrated via measured UE received power in a section of the geographical area. 
Two approaches are formulated to identify candidate RIS locations: reflection-based and scattering-based. Then, the impact is gauged of RIS aperture sizes, number of elements, and RIS density across frequencies. Based on these proxies for cost, light is shed on the tradeoff between the performance enhancement and the capital expense of equipping a network with RIS units.

The main contributions can be distilled as follows.

\begin{itemize}
   
\item 
A ray-based methodology is developed for joint optimization of RIS placement, phase-shift configuration, and BS beamforming. The method incorporates RIS orientation constraints and considers building geometry.

\item 
The impact of design parameters (e.g., RIS aperture, number of elements, and RIS density) on coverage across multiple frequency bands is evaluated in an urban setting.

\item 
Leveraging measured UE data from a production network, site-specific material properties are calibrated. The performance of the RIS placement algorithms is evaluated in both calibrated and uncalibrated conditions.

\item The tradeoff between performance gains and deployment cost is shown to be heavily skewed toward the latter: even under optimistic assumptions about RIS apertures and deployment locations, achieving meaningful improvements in dense urban areas requires a massive rollout.
\end{itemize}

The remainder of the paper is organized as follows. Section~\ref{sec:Sec2} introduces the models, including the cellular network layout, antenna configurations, and ray tracing parameters. Section~\ref{sec:Sec3} presents the proposed data-driven RIS deployment framework, detailing the RIS configuration, and candidate location selection methods. Section~\ref{sec:simulations} provides simulation results to evaluate the impact of the propounded strategies across multiple radio deployments. Section~\ref{sec:calibration} validates the results using material-calibrated ray tracing based on empirical measurements. Section~\ref{sec:conclusion} concludes the paper.


\emph{Reproducible Research.}
To foster reproducibility and further research, the full simulation framework and RIS deployment algorithms are provided as open-source code.\footnote{Available at: \url{https://github.com/Telefonica-Scientific-Research/DDRD}}

\section{Network, RIS, and Propagation Models}
\label{sec:Sec2}


\subsection{Cellular Network Model}

\begin{figure}[t]
    \centering
    \includegraphics[width=\columnwidth]{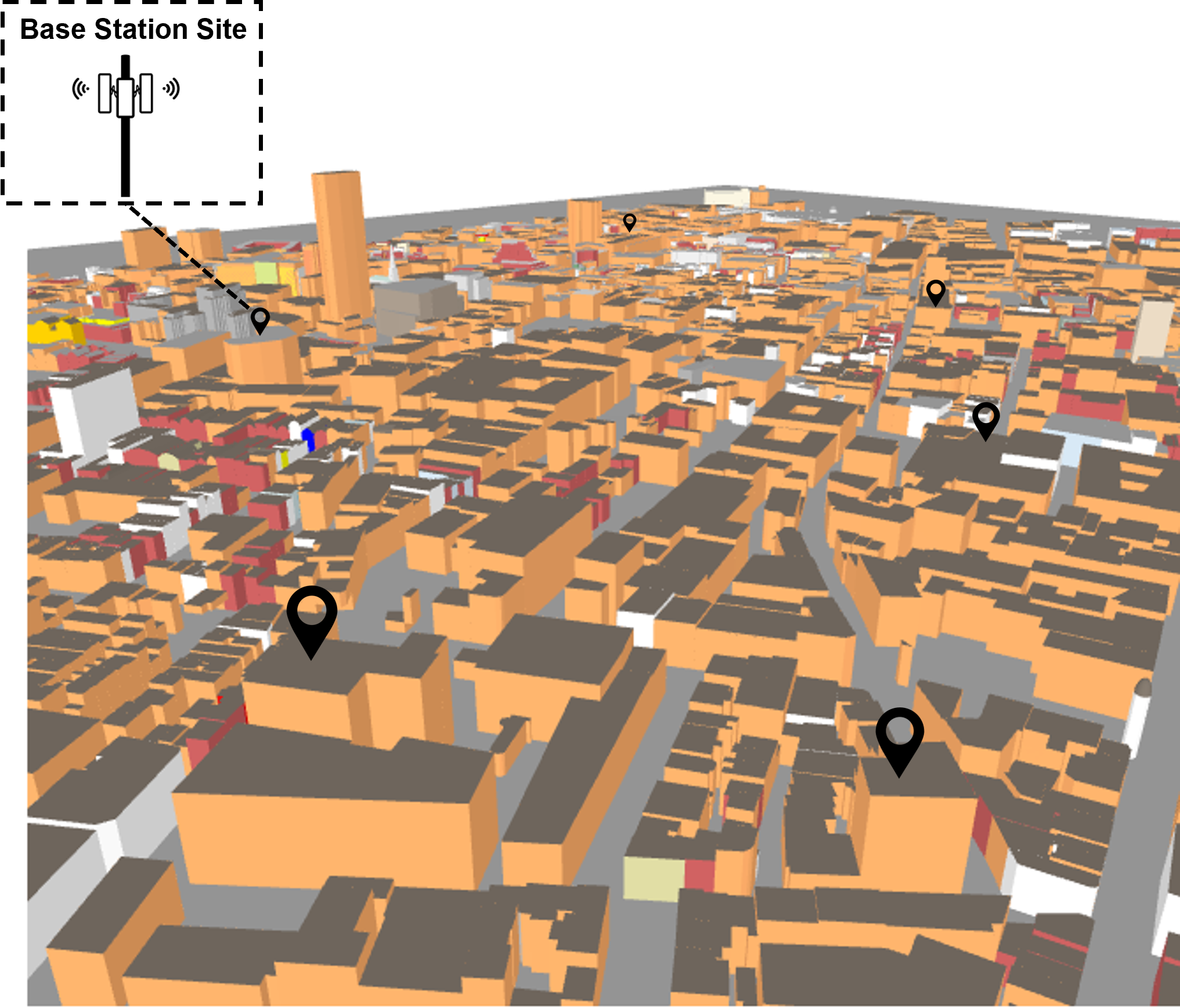}
    \caption{3D visualization of the area under consideration, produced with OpenStreetMap. Black pins indicate BS sites.}
    \label{fig:3D_visualization}
\end{figure}

A commercial cellular network deployed by a leading MNO is considered. 
The studied network segment consists of 12 BSs, with heights ranging from 18 to 56 m. Each BS is equipped with three sector antennas, for a total of 36 cells. This segment spans $1340\,\text{m} \times 1390\,\text{m}$ in the UK. Figure~\ref{fig:3D_visualization} presents a 3D visualization of the area, obtained via OpenStreetMap \cite{OpenStreetMap}, with the locations of several BSs explicitly indicated. 

\subsubsection{Antennas} 

Each BS employs a planar array featuring \( M = M_{\text{h}}M_{\text{v}} \) vertically polarized elements arranged in \( M_{\text{h}} \) columns and \( M_{\text{v}} \) rows. The configuration of the $b$th BS is described by its vertical tilt angle, $\theta^{b}$, and its horizontal bearing angle, $\phi^{b}$, both configured by the MNO.
Each antenna element is supported by a single radio-frequency chain and it abides by the 3GPP radiation pattern, with half-power beamwidths of $\phi^{b}_{\text{3dB}}=65^{\circ}$ and $\theta^{b}_{\text{3dB}}=10^{\circ}$~\cite{3GPP38901}. 

\subsubsection{Radio Deployments} 

The performance is to be evaluated for three distinct systems:
\begin{itemize}
    \item 4G at 2\,GHz (FR1).
    \item 5G at 3.5\,GHz (FR1).
    \item 6G at 10\,GHz (FR3).
\end{itemize}

\noindent The three deployments represent different generations of cellular networks with their typical conditions. The 4G system at 2 GHz (FR1) provides wide-area coverage with strong building penetration and stable outdoor service. The 5G system at 3.5 GHz (FR1) offers higher bandwidth and capacity while still keeping moderate coverage and acceptable penetration for urban areas. The 6G system at 10 GHz (FR3) targets very high data rates with large bandwidth but faces stronger path loss and more blockage, requiring careful site planning~\cite{nr2019base, haneda20165g, ratul2023atmospheric}. Together, these three cases give a realistic view of current and future radio deployments. For each, the parameters (adopted from commercial products, either existing or under development) are summarized in Table~\ref{tab:network_characteristics}. Note that a single polarization is considered, hence the element counts are half what they would be with dual polarization.
The ability to reuse the existing site grid is critical for the economic feasibility of incorporating new spectrum, as additional sites would impose major cost burdens and prolong deployment timelines~\cite{holma2021extreme}. Therefore, the same BS coordinates are considered for the three systems.\footnote{The 6G system operates at a higher frequency and with more bandwidth than its 4G/5G counterparts, but it is configured with a lower transmit power. This reflects a role for 6G as a non-standalone capacity-boosting layer with possibly discontinuous coverage.}
To enable a fair comparison, the three deployments reuse the same BS site grid while differing only in frequency, bandwidth, and transmit power according to commercial specifications for 4G, 5G, and 6G. This isolates the effect of operating band and system configuration, ensuring that the evaluation reflects realistic evolution across cellular generations. While higher-frequency FR3 (7–24 GHz) deployments would typically require denser site grids to compensate for increased path loss, here we intentionally maintain the same grid to assess whether RIS deployment can serve as an alternative to such densification. The 6G system operates at a higher frequency and with more bandwidth than its 4G/5G counterparts, but it is configured with a lower transmit power. This reflects a role for 6G as a non-standalone capacity-boosting layer with possibly discontinuous coverage. A key question addressed in this paper is whether RIS assistance can extend such 6G coverage for outdoor UEs.

\begin{table}[ht]
    \centering
    \caption{System features \cite{lopez2024capacity}}
    \label{tab:network_characteristics}
    \footnotesize 
    \renewcommand{\arraystretch}{1.2} 
    \resizebox{\columnwidth}{!}{ 
    \begin{tabular}{|l|c|c|c|}
        \hline
        \textbf{Feature} & \textbf{4G} & \textbf{5G} & \textbf{6G} \\
        \hline
        Carrier frequency [GHz] & 2 & 3.5 & 10 \\
        \hline
        Bandwidth [MHz] & 20 & 100 & 200 \\
        \hline
        \textbf{Sectors per site} & \multicolumn{3}{c|}{3} \\
        \hline
        \textbf{Polarization} & \multicolumn{3}{c|}{Vertical} \\
        \hline
        Planar array topology & 2×2 & 4×8 & 4×16 \\
        \hline
        Beamforming codebook size & 4 & 32 & 64 \\
        \hline
        TX power per cell [dBm] & 43 & 49 & 44 \\
        \hline
        TX power per subcarrier [dBm] & 12.2 & 13.85 & 8.85 \\
        \hline
        Number of subcarriers & 1200 & 3276 & 3276 \\
        \hline
    \end{tabular}
    } 
\end{table}

\subsubsection{Beamforming Codebook} 

Each BS employs a two-dimensional discrete Fourier transform (DFT) beamforming codebook per system. This codebook defines a set of orthogonal beams on specific angular directions; each such beam is described by the Kronecker product of two one-dimensional DFT vectors, one for azimuth and one for elevation, namely
\begin{equation}
    \boldsymbol{w}_{m_{\text{h}}, m_{\text{v}}} = \boldsymbol{w}_{m_{\text{h}}} \otimes \boldsymbol{w}_{m_{\text{v}}},
    \label{eq:DFT_UPA}
\end{equation}
where \( m_{\text{h}}, m_{\text{v}} \) are the horizontal and vertical beam indices. The horizontal DFT vector is given by
\begin{equation}
\boldsymbol{w}_{m_{\text{h}}} = \frac{1}{\sqrt{M_{\rm h}}}
  \left[1,\,e^{-j\frac{2\pi m_{\rm h}}{M_{\text{h}}}},\,
        \cdots,\,e^{-j\frac{2\pi m_{\rm h}(M_{\rm h}-1)}{M_{\rm h}}}\right]^{\text{T}},
\label{eq:DFT_ULA}
\end{equation}
for $m_{\text{h}} = 0,1,\ldots,M_{\text{h}}-1$. The vertical DFT vector follows the same formulation.
As a result, each BS can select a beam with elevation and azimuth angles satisfying
\begin{equation}
    \sin \theta_{m_{\text{v}}} = \frac{2 \, m_{\text{v}}}{M_{\text{v}}} - 1 \qquad\quad \sin \phi_{m_{\text{h}}} = \frac{2 \, m_{\text{h}}}{M_{\text{h}}} - 1.
    \label{eq:beam_steering}
\end{equation}

\subsubsection{UEs}

UEs are deployed on a regular grid that partitions the area into tiles of 2\,m $\times$ 2\,m at a height of 1.5\,m. The received power is averaged over each tile. The focus is on outdoor UEs, as outdoor coverage remains the primary use-case for cellular networks, yet the approach can be extended to indoor environments provided a floor plan is available and the ray-tracing tool supports outdoor-to-indoor propagation. 
Each UE is equipped with a single isotropic antenna having vertical polarization. 
A single UE is served on each time-frequency signaling resource, isolating the effects of scheduling. 

\subsection{ Physically consistent RIS Model}
\label{sec:RISmodel}

To overcome the RIS modeling limitations detailed in Sec.~\ref{sec:Sec1}, the physically consistent representation introduced in~\cite{9713744} is adopted, which treats the RIS as a continuous reradiating surface rather than a discrete array of independent elements. This approach captures reradiation effects by applying a spatial modulation function derived from ray optics~\cite{yablonovitch1982statistical}.
Precisely, the spatial modulation function \( \Gamma(x, y) \) describes the complex transformation applied by the RIS at each surface point, combining amplitude and phase modulation into \cite{10419169}
\begin{equation}
    {\Gamma}(x, y) = R \sqrt{\eta} {A}(x, y) e^{ j \, {\varphi}(x, y) },
\end{equation}
where \( (x, y) \) is a local coordinate on the RIS, \( R \) accounts for the effect of small-scale surface roughness on the RIS, the surface efficiency \( \eta \) denotes the fraction of incident power effectively reradiated, and \({A}(x, y) \) is the modulation amplitude which controls how much of the incident power is reradiated at each point on the RIS \cite{ShaSimGer2024}sc.

This model enables physically consistent simulations while accounting for finite surface size, angular reradiation constraints, and energy conservation. The modulation amplitude is not the local reflection coefficient 
assumed in the diagonal matrix approach \cite{10473672}.
Rather, it is a macroscopic spatial modulation function that relates the amplitude of the scattered waves to the incident field, while accounting for finite surface size, angular reradiation constraints, energy conservation, and boundary conditions. This physically consistent ray-based modeling of RIS is implemented in Sionna RT, and it is leveraged in our study.

\subsection{Site-specific Ray Tracing and Coverage Calculation}

The propagation among BSs, UEs, and RIS units is modeled using Sionna RT~\cite{10705152}, an open-source ray-tracing simulator developed by NVIDIA. It enables a physically accurate representation of electromagnetic wave propagation, capturing reflections, diffraction, scattering, and multipath. Balancing realism and computational efficiency, the ray-tracing engine is configured with the parameters in Table~\ref{table: ray parameters} unless otherwise stated. A specular reflection ray obeys Snell’s law and results from a mirror-like bounce off a surface. A scattering ray, on the other hand, arises from a non-specular interaction that spreads energy across a range of directions—typically caused by rough surfaces or small-scale irregularities. These interaction types are explicitly and distinctly modeled during ray tracing in Sionna~\cite{10705152}, enabling separate treatment of their respective physical effects, and in Sections~\ref{sec:reflection_based_approach} and~\ref{sec:scattering_based_approach}, we show that they play key roles in RIS deployment.

To compute the large-scale channel gain between each BS and UE, the set of corresponding rays is considered. 
The geographical region of interest is partitioned into square tiles \( C_{p,q} \) of size 2\,m $\times$ 2\,m, where each tile corresponds to a UE. For each BS sector \( t \) and DFT beam index \( m  \), the directional channel gain at tile \((p,q)\) is computed as
\begin{equation}
    G_{t,m,p,q} = \frac{1}{|C_{p,q}|}  \iint_{C_{p,q}} \left| \boldsymbol{h}_{t,p,q}^{\text{H}}(x,y) \boldsymbol{w}_{t,m} \right|^2 \, dx dy,
    \label{eq:path_gain}
\end{equation}
where \( \boldsymbol{h}_{t,p,q}(x,y) \) is the channel vector at position \( (x,y) \), obtained by all ray types (LoS, reflected, diffracted, and scattered), \( \boldsymbol{w}_{t,m} \) is the DFT precoding vector, and \( |C_{p,q}|\) is the tile area. The vector $\boldsymbol{h}$ in (6) contains the complex path coefficients generated by the Sionna RT engine. When no RIS is present, $\boldsymbol{h}$ only captures the BS--UE propagation paths (including LoS, reflections, diffractions, and scattering). When a RIS is deployed, $\boldsymbol{h}$ also includes the Tx--RIS--Rx paths, in which the reradiation by the RIS is modeled through the spatial modulation function \( \Gamma(x, y) \) defined in (5). Thus, the effect of \( \Gamma(x, y) \) is embedded in the path coefficients $\boldsymbol{h}$, and (6) directly computes the corresponding channel gain based on these coefficients.

The reference signal received power (RSRP) for tile \( (p,q) \), served by BS sector \( t \) and beam index \( m \), is then 
\begin{equation}
    \text{RSRP}_{p,q} = G_{t,m,p,q} P_{{\text{t}}},
    \label{RSS_computation}
\end{equation}
where \( P_{\text{t}} \) is the transmit power per subcarrier, as specified in Table~\ref{tab:network_characteristics}. The UE at the center of each tile selects its serving BS and beam by  choosing the maximum RSRP value across all candidate BSs, sectors, 
and beam indices. This corresponds to the standard UE association 
rule in cellular networks, whereby each UE connects to the BS providing 
the strongest received power, rather than summing contributions from 
multiple BSs. We use RSRP as the outage metric since it is the industry-standard indicator of cellular coverage, consistently adopted in both technical practice and regulatory frameworks~\cite{cablefree_rsrp,comreg2021}.

\begin{table}[b]
\centering
\caption{Ray-tracing parameters}
\label{table: ray parameters}
\def\arraystretch{1.2}
\begin{tabulary}{\columnwidth}{ |p{2.8cm} | p{4.8cm} | }
\hline
\textbf{Parameter}	& \textbf{Value}\\ \hline
Method 			& Fibonacci, shoot-and-bounce approach\\ \hline
  Ray types			& Specular reflection, diffraction, scattering \\ \hline
  Number of shooting rays 		&  ${10^7}$ per cell\\ \hline
  	Maximum bounces 				& $4$ per ray \\ \hline
    Grid resolution 			& 2\,m × 2\,m\\ \hline
    Number of BSs 			& 36 (12 sites, 3 sectors per site)\\ \hline

\end{tabulary}
\end{table}

\begin{figure*}[t]
    \centering
    \subfloat[4G deployment at 2\,GHz]{%
        \includegraphics[width=0.32\textwidth]{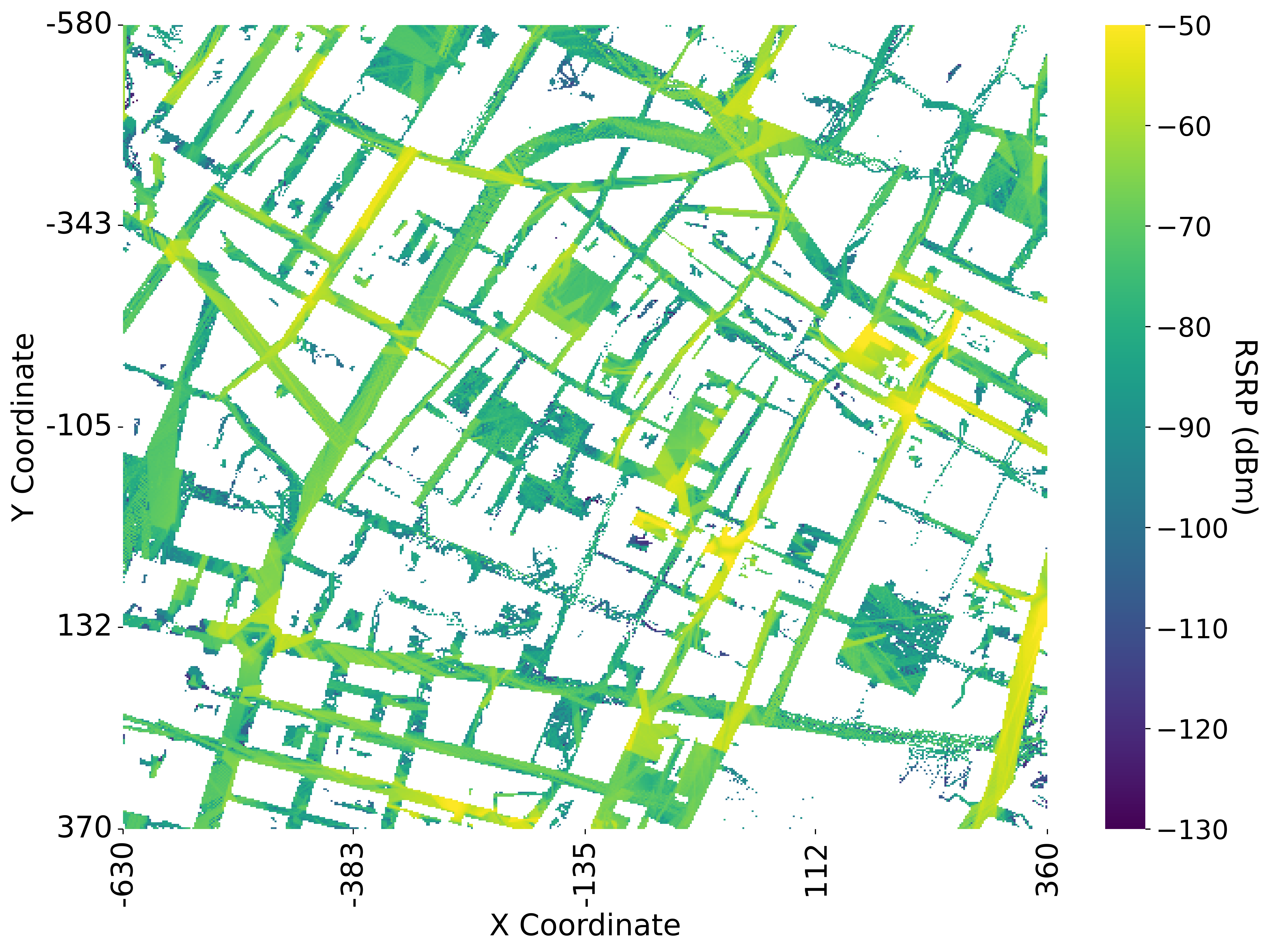}
        \label{fig:Heatmap_2GHz}
    }
    \hfill
    \subfloat[5G deployment at 3.5\,GHz]{%
        \includegraphics[width=0.32\textwidth]{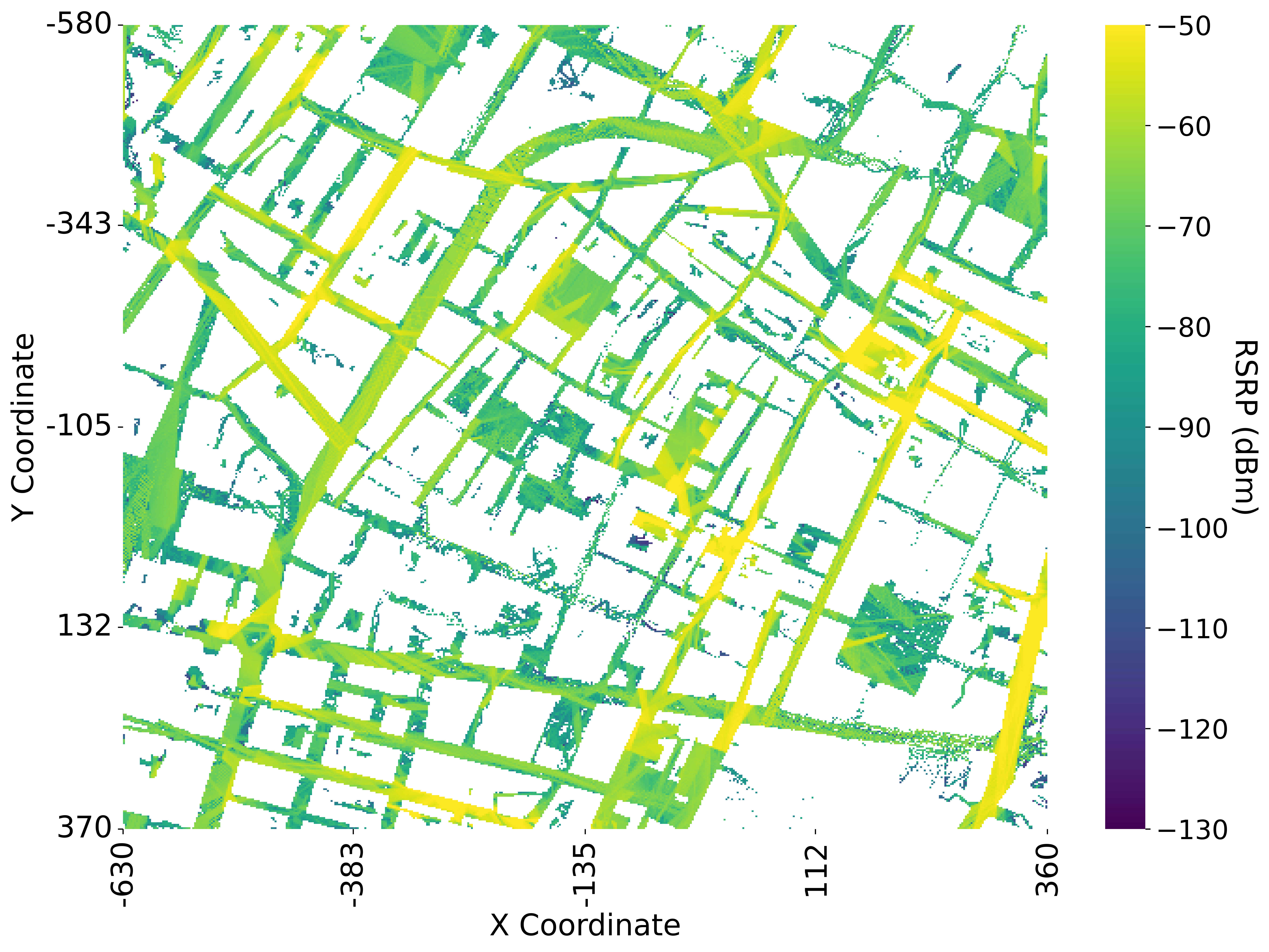}
        \label{fig:Heatmap_3_5GHz}
    }
    \hfill
    \subfloat[6G deployment at 10\,GHz]{%
        \includegraphics[width=0.32\textwidth]{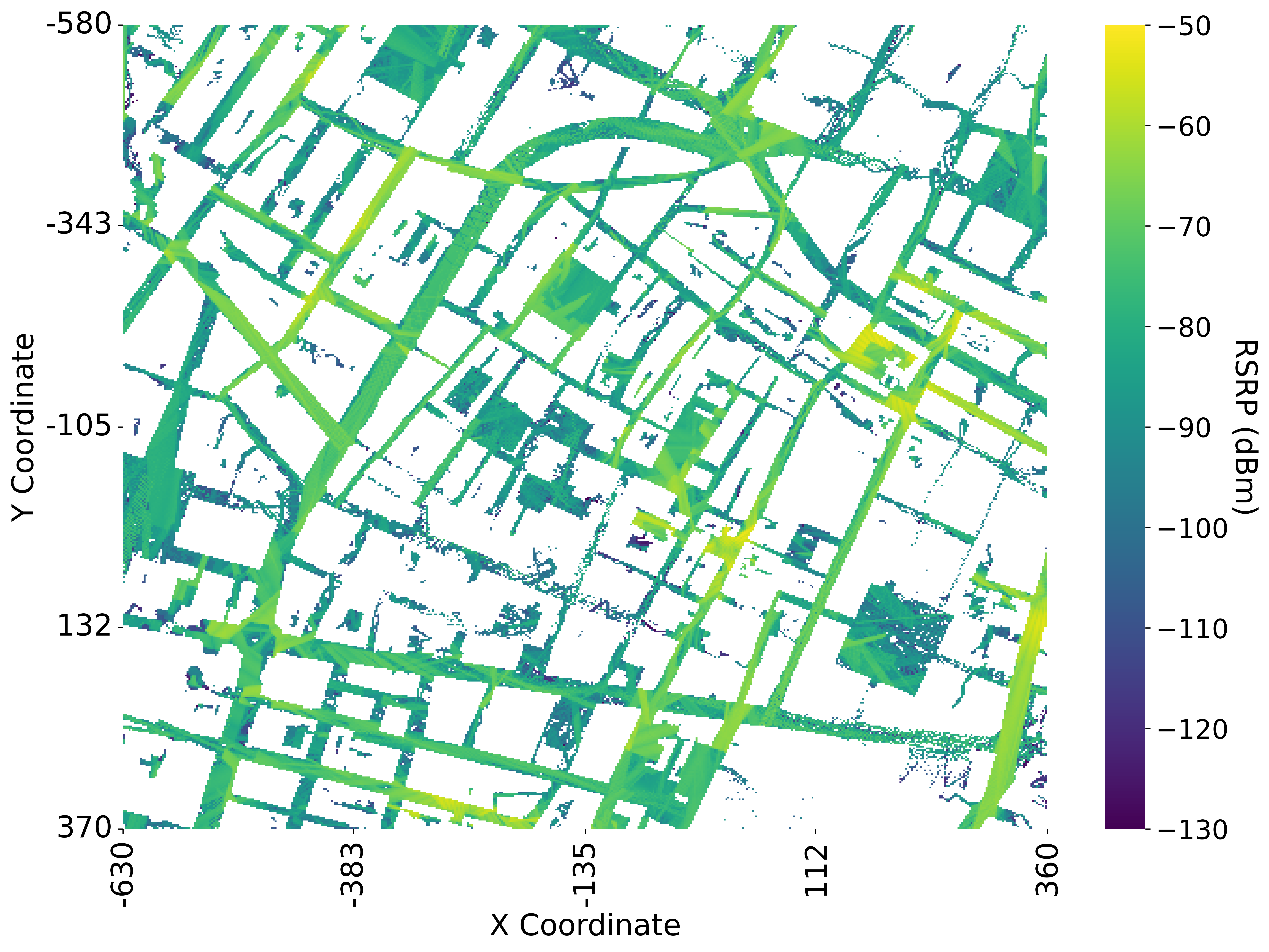}
        \label{fig:Heatmap_10GHz}
    }
    \caption{RSRP heatmaps across the considered urban area for 4G, 5G, and 6G.}
    \label{Heatmap}
\end{figure*}

\section{Data-Driven RIS Deployment}
\label{sec:Sec3}

This section presents the data-driven strategy for large-scale RIS deployment.

\begin{figure}[t]
    \centering
    \includegraphics[width=0.88\columnwidth]{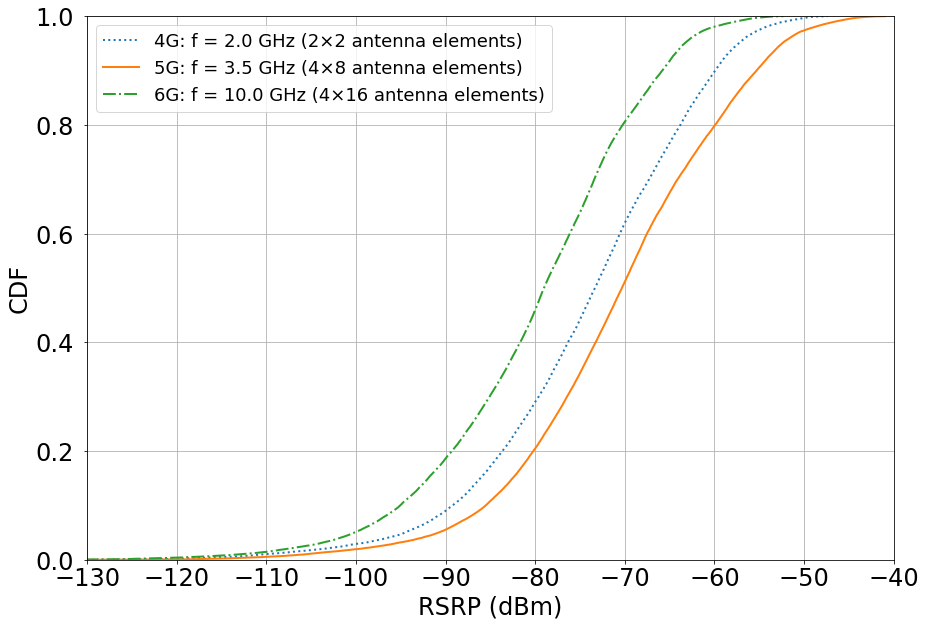}
    \caption{RSRP distribution before RIS deployment for 4G (blue), 5G (orange), and 6G (green).}
    \label{fig:CDF_RSS_Network}
\end{figure}

\subsection{Coverage Evaluation and Outage Detection}
\label{subsec:coverage_evaluation}

\begin{figure}[t]
    \centering
    \includegraphics[width=0.85\columnwidth]{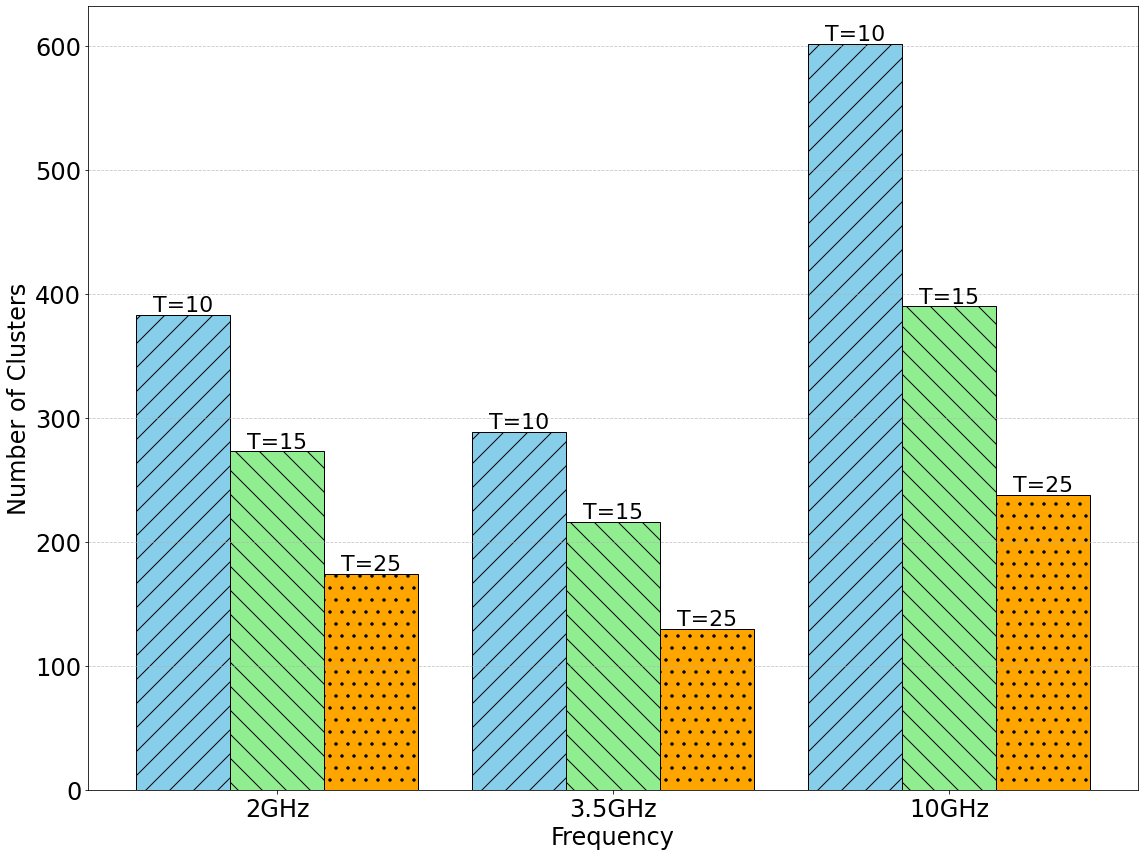}
    \caption{Number of clusters formed by the BIRCH algorithm vs. parameter \( T \) across the considered frequency bands.}
    \label{fig:number_clusters_vs_freq}
\end{figure}

The baseline network coverage is evaluated by deploying BSs and distributing UEs on a grid, as described in Sec.~\ref{sec:Sec2}.
Figure~\ref{Heatmap} illustrates the RSRP heatmaps for the considered urban area. UEs with an RSRP below -100\,dBm are classified as \textit{outage UEs} \cite{teltonika}. Based on this threshold, the outage percentages are 2.35\%, 1.86\%, and 5.15\% at 2\,GHz, 3.5\,GHz, and 10\,GHz, respectively.
As expected, coverage deteriorates at 10\,GHz 
due to increased path and penetration losses. 
To further assess the RSRP across the network, Fig.~\ref{fig:CDF_RSS_Network} shows its cumulative distribution function (CDF), which evidences how the RSRP in the three systems is also influenced by the beamforming gain. 
The goal of a strategic RIS deployment is to enhance the RSRP at outage locations. However, given the large number of outage UEs and their spatial distribution, deploying a separate RIS for each UE is unfeasible.

\begin{figure*}[t]
    \centering
    \subfloat[4G deployment at 2\,GHz]{%
        \includegraphics[width=0.32\textwidth]{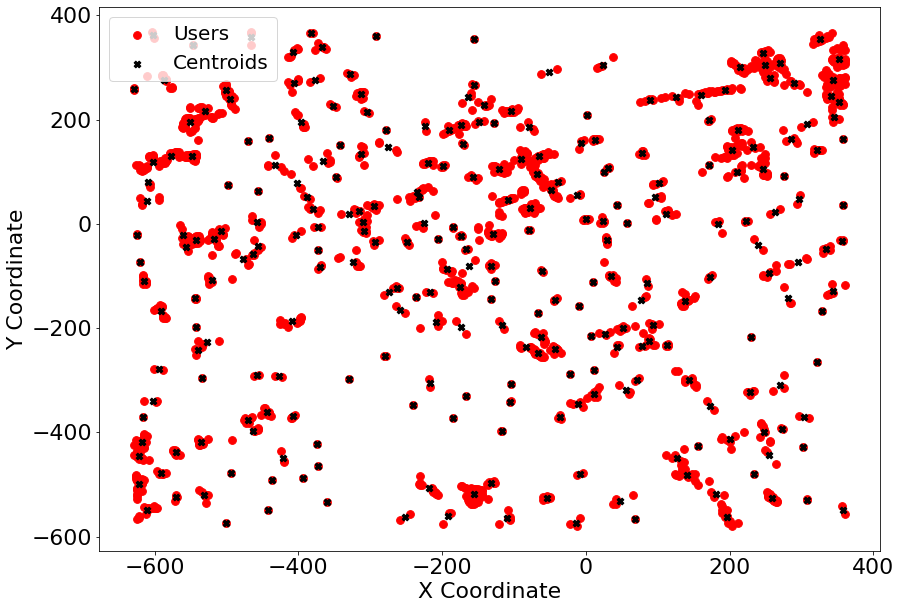}
        \label{fig:cluster_2}
    }
    \hfill
    \subfloat[5G deployment at 3.5\,GHz]{%
        \includegraphics[width=0.32\textwidth]{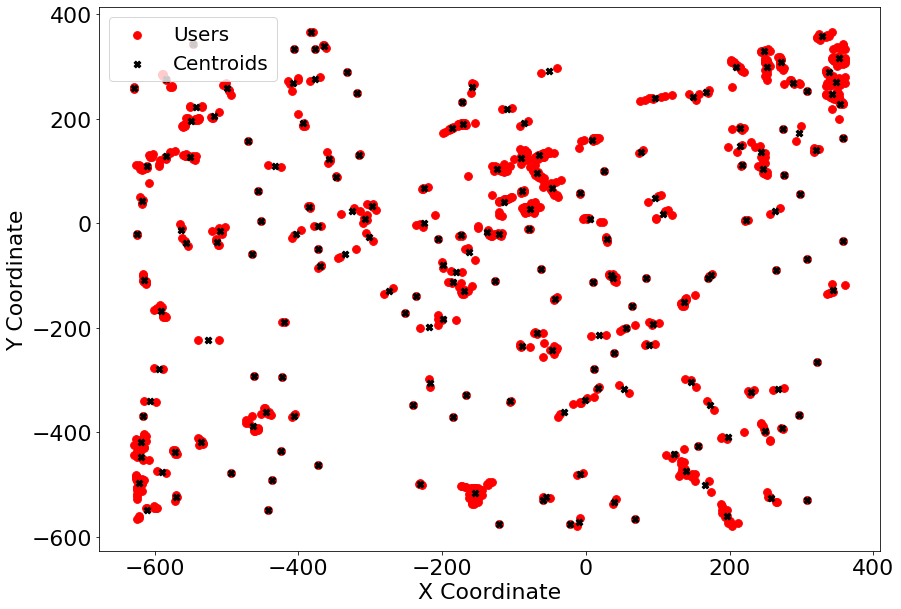}
        \label{fig:cluster_3_5}
    }
    \hfill
    \subfloat[6G deployment at 10\,GHz]{%
        \includegraphics[width=0.32\textwidth]{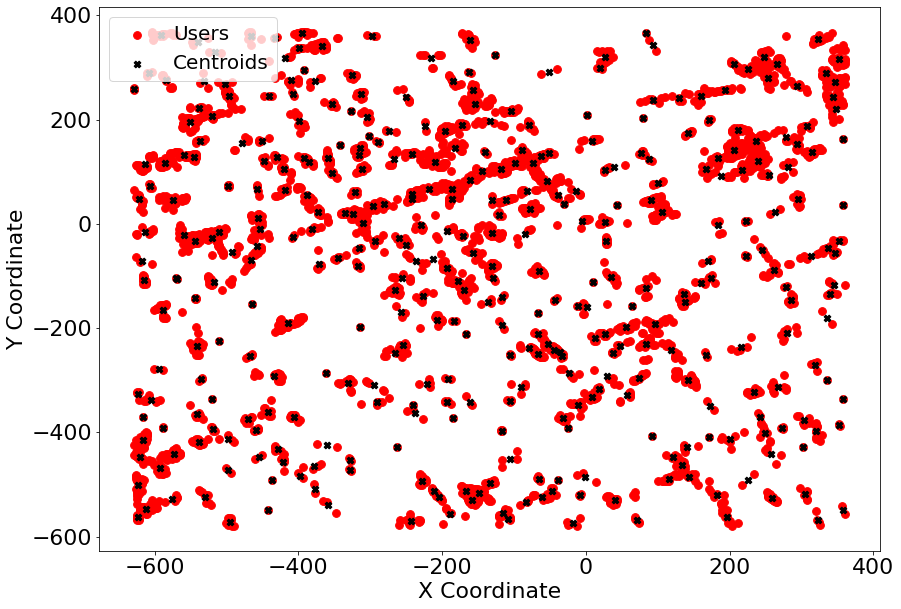}
        \label{fig:cluster_10}
    }
    \caption{Scatter plot of outage UEs and their clustering.
    }
    \label{fig:clustering}
\end{figure*}

\subsection{Clustering of Outage Locations}
\label{subsec:clustering}

To address the scalability challenge posed by the large number of outage UEs, adjacent ones are grouped into clusters; then,
each RIS can serve a cluster rather than an individual UE.

To group outage UEs efficiently, we employ the balanced iterative reducing and clustering using hierarchies (BIRCH) algorithm~\cite{10.1145/233269.233324}. BIRCH is a scalable hierarchical clustering method that incrementally builds compact clustering feature (CF) trees, enabling fast and memory-efficient clustering of large datasets without requiring the number of clusters to be predefined. Each cluster is represented by a CF vector
\begin{equation}
\mathrm{CF} = \left[ U, \sum_{i=1}^{U} \boldsymbol{s}_i, \sum_{i=1}^{U} \|\boldsymbol{s}_i\|^2 \right],
\label{eq:cluster_feature}
\end{equation}
where \( U \) is the number of UEs and \( \boldsymbol{s}_i \in \mathbb{R}^2 \) denotes the 2D position \((x, y)\) of UE \( i \) (center of a tile) in a Cartesian coordinate system.

Clusters are merged based on a threshold distance \( T \) that controls their compactness and determines the number of necessary RIS. Compared to K-means~\cite{1056489}, BIRCH does not require predefining the number of clusters and can incrementally adapt to the spatial distribution of outage UEs, which is critical for large-scale, site-specific deployments. Additionally, K-Means requires the number of clusters (K) to be fixed a priori, is sensitive to centroid initialization, and has the limitation of creating spherical clusters due to the distance computation with respect to the centroid of each cluster, leading to unstable or suboptimal results in realistic wireless environments. Unlike DBSCAN~\cite{10.5555/3001460.3001507}, BIRCH is less sensitive to non-uniform cluster densities and scales more efficiently when handling tens of thousands of UEs.

The threshold \(T\) is a key design parameter that defines the maximum allowable Euclidean distance (in meters) between data points within a cluster in the BIRCH algorithm. A smaller value results in a larger number of clusters (and thus of RIS), while a larger value reduces the number of RIS at the cost of an increased coverage area for each. In this work, the selected value \(T = 15\)~m was determined through extensive empirical testing to balance clustering granularity, RIS coverage capability, and deployment practicality in dense urban scenarios. This value ensures that a single RIS can effectively serve all UEs within its cluster while avoiding excessive deployment density. Figure~\ref{fig:number_clusters_vs_freq} shows how the number of clusters varies with \(T\) across the considered frequencies. In turn, Figure~\ref{fig:clustering} illustrates the clustering for outage UEs using \(T = 15\)~m, with cluster centroids indicated by cross symbols.

\subsection{RIS Candidate Locations: Reflection-based Approach}
\label{sec:reflection_based_approach}

For each cluster of outage UEs determined through BIRCH, one RIS unit is deployed to maximally improve coverage, either by aligning with the reflection point of the strongest ray or by leveraging the first and second reflection points of alternative rays. Here, “alternative rays” denote the weaker specular-reflection rays, beyond the strongest one, that still reach the receiver and can provide additional candidate points for RIS deployment.
This study assumes that a RIS can be deployed on any building surface where a valid reflection or scattering point exists, providing an upper bound on performance as deployment constraints such as physical accessibility or regulations are neglected.

This section presents two algorithms for reflection-based identification of candidate RIS locations:
\begin{itemize}
    \item \textit{Strongest ray selection:} Uses the highest-power reflection ray received at each cluster centroid. The strongest ray refers to the full multi-bounce propagation path that yields the maximum received power. All reflection points along this path are tested as potential RIS locations.
    \item \textit{All-ray selection:} Considers the first and second reflection points of all received rays at each cluster centroid. These reflection points are collected across all rays, and involve different surfaces than those found along the strongest ray path. This broader selection allows the algorithm to identify additional candidate RIS locations that may be effective when assisted by reradiation.
\end{itemize}


The central idea in both aproaches is that reflection points along a ray's path are desirable RIS deployment locations. By considering the rays received at each cluster's centroid, as a proxy for the UEs in the cluster, one or more candidate locations are identified from which energy can be effectively redirected to that cluster. 

\subsubsection{Strongest Ray Selection}

Consider the centroid of the $k$th cluster, which receives a set of rays. 
The RIS is tested at each reflection point of the strongest ray, with its phase profile configured as detailed in Sec.~\ref{sec:RIS_config_beamforming}.
For each such point, the RSRP with RIS assistance at the centroid tile is obtained and denoted by $P^{\text{RIS}}_k$, and the RIS position that yields the greatest improvement over the baseline value (without RIS) is selected. Then, the algorithm proceeds to evaluate whether the selected RIS placement substantially improves the coverage for the cluster. If more than 40\% of UEs show an RSRP improvement, the cluster is marked as RIS-effective. 
This 40\% threshold was determined through extensive trial-and-error simulations to provide a practical balance between RIS coverage gain and computational cost: higher thresholds increase the computational load with only marginal additional gain, while lower thresholds reduce the processing time but deliver less improvement.  
Clusters that do not meet this criterion are deferred for further optimization by executing the all-ray selection algorithm (see below). The entire procedure is detailed in Alg.~\ref{alg:RIS_location_optimization}.

Clusters where the foregoing algorithm fails to improve the performance typically exhibit one or several of the following:
\begin{itemize}
    \item Blocked or ineffective strongest-ray reflection locations (e.g., obstructed LoS between RIS and UEs).
    \item Centroids that are poorly representative of the spatial distribution of UEs.
    \item Reflection points located at low elevations, limiting RIS reradiation effectiveness.
\end{itemize}

\begin{algorithm}[t]
\caption{Strongest Ray Selection.}
\label{alg:RIS_location_optimization}

\SetKwInOut{Input}{Input}
\SetKwInOut{Output}{Output}

\Input{Centroid of each cluster, set of UEs $\mathcal{U}_k$}
\Output{Optimized RIS location for each cluster}

\For{each centroid }{
    Identify the set of received rays;
    
    Determine strongest ray;

    Extract reflection locations $\{L_k\}$\;
    
    Initialize $P_{\max, k}^{\text{RIS}} = 0$\;

    \For{$k = 1$ to $4$ in $\{L_k\}$}{
        Deploy RIS at $L_k$\;
        
        Run BS beamforming optimization\;

        Run RIS phase shift configuration\;
        
        Compute $P_{{k}}^{\text{RIS}}$\;
        
        \If{$P_{k}^{\text{RIS}} > P_{\max, k}^{\text{RIS}}$}{
            Update $P_{\max, k}^{\text{RIS}} \gets P_{k}^{\text{RIS}}$\;
            Update $L_{\text{RIS}} \gets L_k$\;
        }
    }
    
    \If{$P_{\max, k}^{\text{RIS}} \leq P_{k}$ }{
        Execute all-ray selection algorithm\;
        \textbf{Continue}\;
    }
    
    Deploy RIS at the optimal location $L_{\text{RIS}}$\;
    
    \For{each user $u \in \mathcal{U}_k$}{
        Align RIS phase shift toward each $u$\;
        
        Compute $P_{u}^{\text{RIS}}$\;
    }
    
    \If{
    $
    \frac{1}{|\mathcal{U}_k|}  \sum_{u \in \mathcal{U}_k} 1 (P_{u}^{\text{RIS}} > P_u)
    > 0.4
    $
    }{
        Classify as a RIS-effective cluster\;
    }
    \Else{
        Execute all-ray selection algorithm\;
    }
}
\end{algorithm}

\subsubsection{All-Ray Selection}

To complement the strongest-ray selection approach, whose weakness is the reliance on a single ray, the all-ray selection considers the first and second reflection points from \textit{all} rays received at the cluster centroid. Physically, the “first and second reflection points” correspond to the first and second bounces of a ray as it propagates through the environment, with the maximum number of bounces limited to four in our ray-tracing configuration. Higher-order reflections (third and fourth) are excluded because they undergo severe attenuation—especially at high frequencies—and would not deliver sufficient power to the RIS. This increases the likelihood of finding effective RIS locations, especially in complex urban settings. 
If multiple candidate points offer similar gains, the one closest in 3D distance to the centroid is selected, to reduce the overall path loss. The algorithm assesses whether the selected RIS location improves the RSRP at centroid tile and subsequently verifies whether more than 40\% of UEs in the cluster benefit. If this condition is met, the RIS deployment is accepted; otherwise:
\begin{itemize}
    \item If some UEs improve and others do not, the non-improved UEs are passed on to the re-clustering process (Sec.~\ref{sec:reclustering}).
    \item If the RSRP of the centroid tile does not improve, the cluster is returned to the re-clustering process  (Sec. \ref{sec:reclustering}).
\end{itemize}

Even with its broader selection, the all-ray algorithm may fail in cases where few or no rays are received at the centroid, or the initial clustering parameter \( T \) groups UEs too loosely, leading to ineffective joint support from a single RIS.


\subsection{RIS Candidate Locations: Scattering-Based Approach}
\label{sec:scattering_based_approach}

As an alternative to reflection-based strategies, a scattering-based method is set forth to identify candidate RIS deployment locations. 
Only single-bounce scattering rays are activated in the ray tracer and the number of ray shootings is increased to $3 \times 10^7$, ensuring dense angular coverage. As a result, the scattering-based algorithm requires execution on a high-performance computing infrastructure.

Among the generated scattering rays that reach a given cluster centroid, the ones retained are those that: (i) reach the centroid after one scattering event, and (ii) form a geometrically valid LoS path from a BS to the centroid via the scatter point. 
Candidate RIS locations are then extracted from the corresponding scatter points. If multiple candidate locations exist, the one closest in 3D distance to the centroid is selected.  

If the centroid's RSRP improves, UE-level gains are evaluated. As in earlier algorithms, if more than 40\% of cluster UEs benefit from the deployment, the RIS is retained; otherwise, the remaining UEs are deferred to the next refinement stage.

The pseudo-code of the scattering-based approach is omitted for the sake of conciseness, as it is structurally similar to Alg.~\ref{alg:RIS_location_optimization}.

\subsection{RIS Configuration and BS Beamforming}
\label{sec:RIS_config_beamforming}


Once a candidate location has been selected, the RIS thereon must be properly oriented, configured, and integrated into the network.

\begin{itemize}


\item Each RIS is positioned along the building's wall at the selected candidate location, making sure it faces outwards.

\item Given a BS location, a UE location, and a RIS location, the spatial modulation function described in Sec.~\ref{sec:RISmodel} is computed.




\item The serving BS must select the beam that maximizes the RSRP at the RIS location. This procedure is iterated over the available DFT beams, and the one with the  highest directional gain is selected.
\end{itemize}


\begin{figure*}[t]
    \centering
    \subfloat[4G at 2\,GHz]{%
        \includegraphics[width=0.32\textwidth]{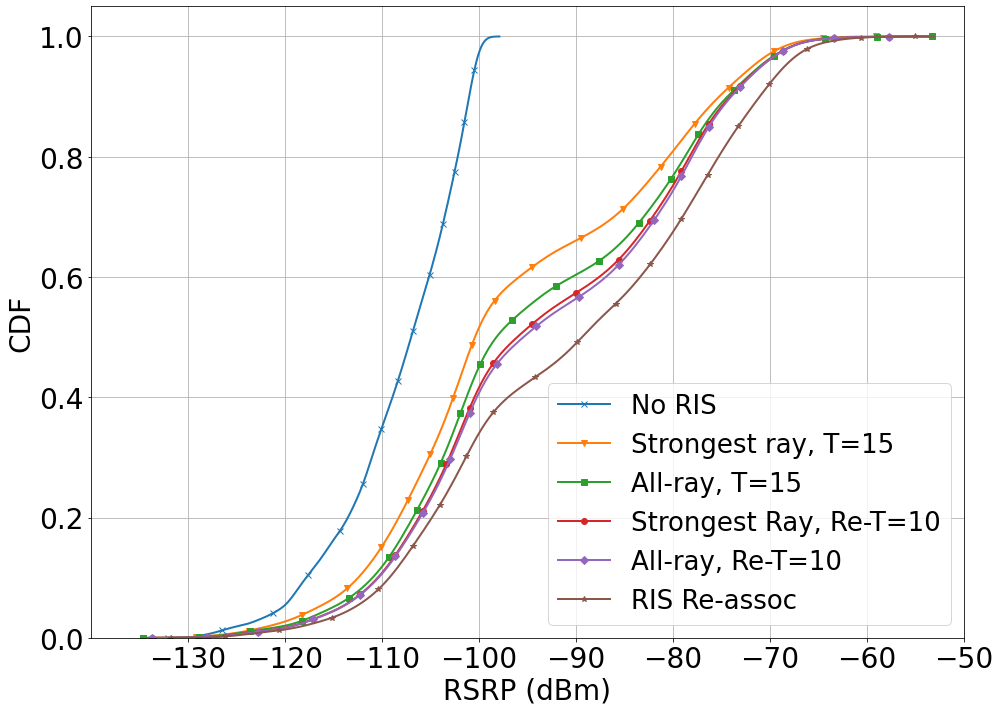}
        \label{fig:Optimization at 2GHz}
    }
    \hfill
    \subfloat[5G at 3.5\,GHz]{%
        \includegraphics[width=0.32\textwidth]{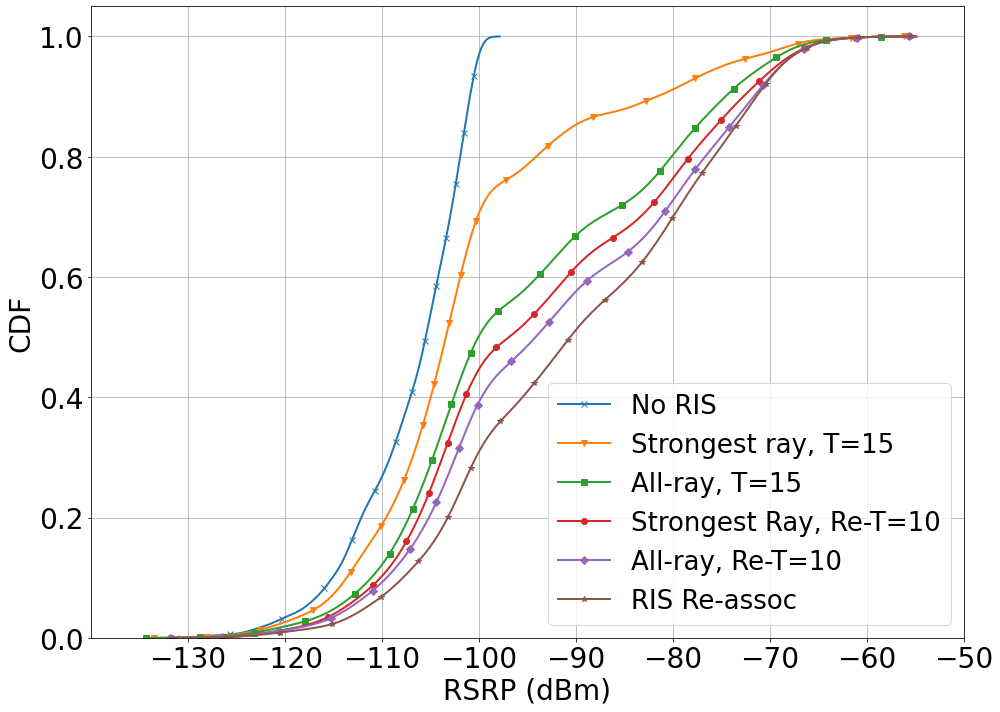}
        \label{fig:Optimization at 3.5GHz}
    }
    \hfill
    \subfloat[6G at 10\,GHz]{%
        \includegraphics[width=0.32\textwidth]{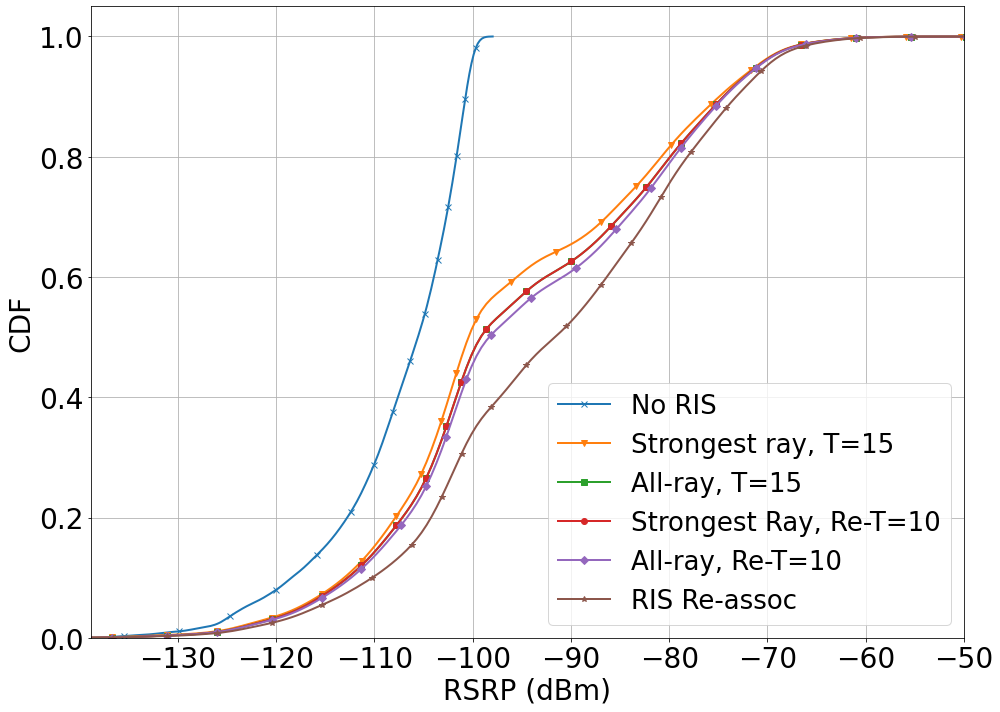}
        \label{fig:Optimization at 10GHz}
    }

    \vspace{0.5em} 

    \caption{RSRP enhancement when deploying the maximum number of RIS following the reflection-based algorithms.}
    \label{fig:CDF_Comparison}
\end{figure*}

\subsection{Re-Clustering and Re-Association}
\label{sec:reclustering}

Let us now address the residual outage UEs for which candidate RIS locations identified by the reflection-based and scattering-based methods were ineffective. Two fallback strategies are employed: re-clustering, and RIS re-association.

\subsubsection{Re-Clustering}

Here, the remaining outage UEs are regrouped using the BIRCH algorithm with a smaller threshold \( T \), yielding more compact and localized clusters. The new threshold
is determined iteratively to optimize the trade-off between coverage and deployment cost.
For each new cluster (i) the centroid location is extracted, and (ii) either the reflection-based or the scattering-based approach is reapplied to find a suitable RIS location.
While reducing $T$ does improve the effectiveness, it also increases the number of RIS units, leading to higher capital costs for the MNO.

\subsubsection{RIS Re-Association}

Some outage UEs may be located in areas where previously deployed RIS units are effective, but were not originally assigned to those UEs' cluster. To avoid unnecessary new deployments, an RIS re-association procedure can assign leftover UEs to existing RISs based on LoS feasibility.
It is a four-step procedure:

\begin{enumerate}
    \item 
    Identify which deployed RISs have a LoS path to each outage UE.
    \item 
    For each candidate RIS, identify which BSs have a LoS path to it.
    \item 
    Among the feasible RISs, select the one closest to the UE.
    \item 
    Among the BSs linked to the selected RIS, choose the closest one.
\end{enumerate}

\noindent This procedure results in an updated association between UEs, RISs, and BSs, improving RSRP without requiring additional RIS deployments.


\section{Simulation Results}
\label{sec:simulations}


\subsection{Performance of Reflection-Based Algorithms}
\label{subsec:per_reflect}

\begin{figure}[t]
    \centering
    \includegraphics[width=\columnwidth]{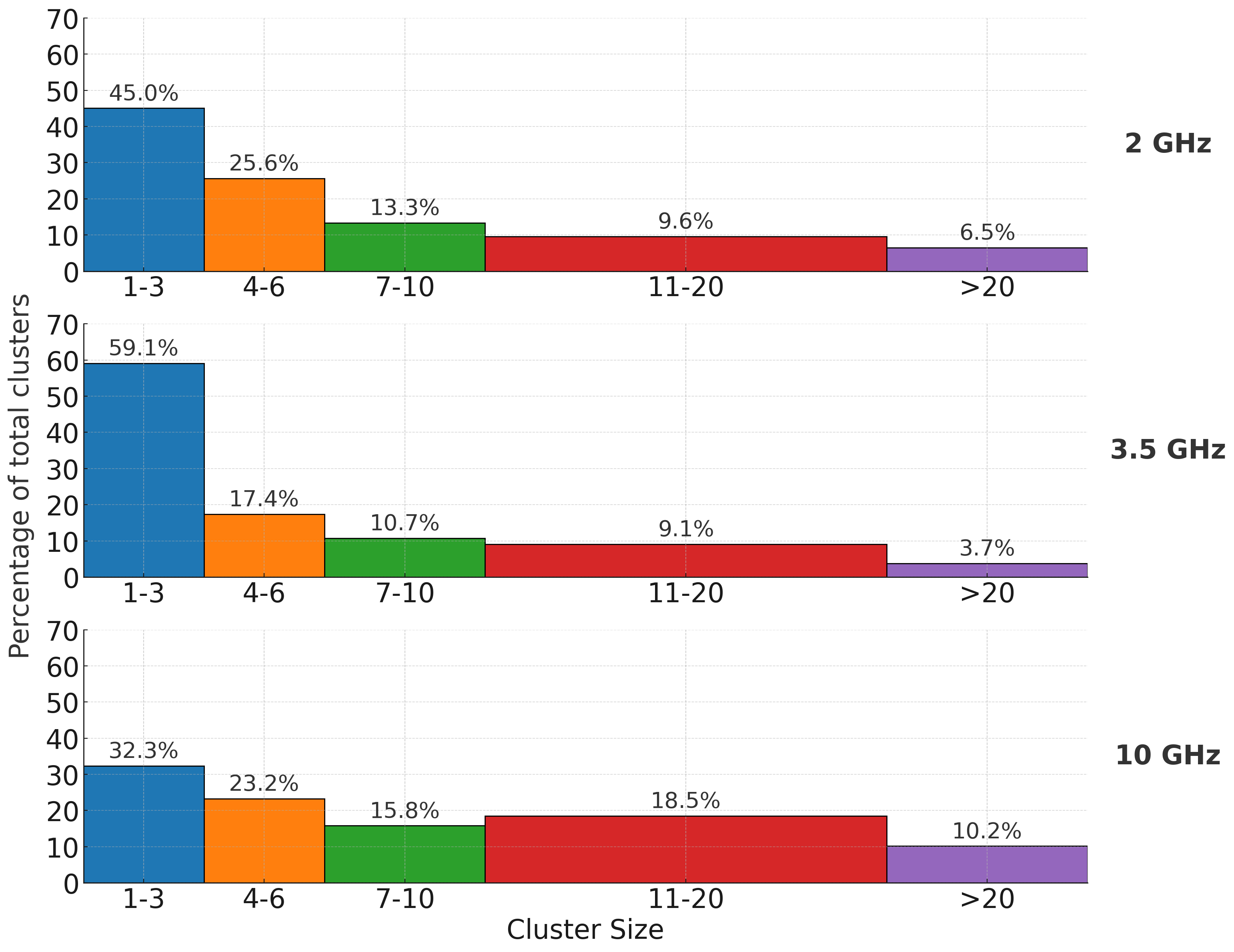}
    \caption{
    Distribution of the number of UEs per cluster at different frequencies; the percentages indicate the ratio of UEs that are within a cluster of the corresponding cluster size.
    }
    \label{fig:user_bar_distribution}
\end{figure}

For an initial test of the reflection-based algorithms, a very large RIS (11.24  $\times$ 11.24 m) is considered with a half-wavelenth element spacing.
Outage UEs are clustered using BIRCH with a threshold of \( T = 15 \) m.

\begin{figure*}[t]
    \centering
    \subfloat[4G at 2\,GHz]{%
        \includegraphics[width=0.32\textwidth]{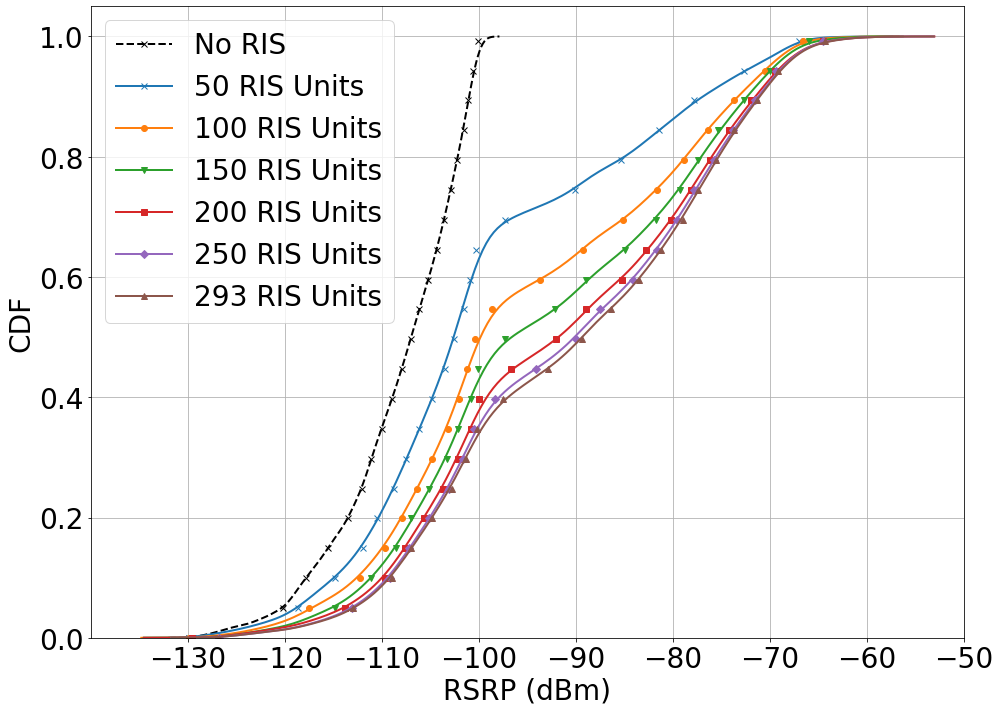}
        \label{fig:inc_RIS_2}
    }
    \hfill
    \subfloat[5G at 3.5\,GHz]{%
        \includegraphics[width=0.32\textwidth]{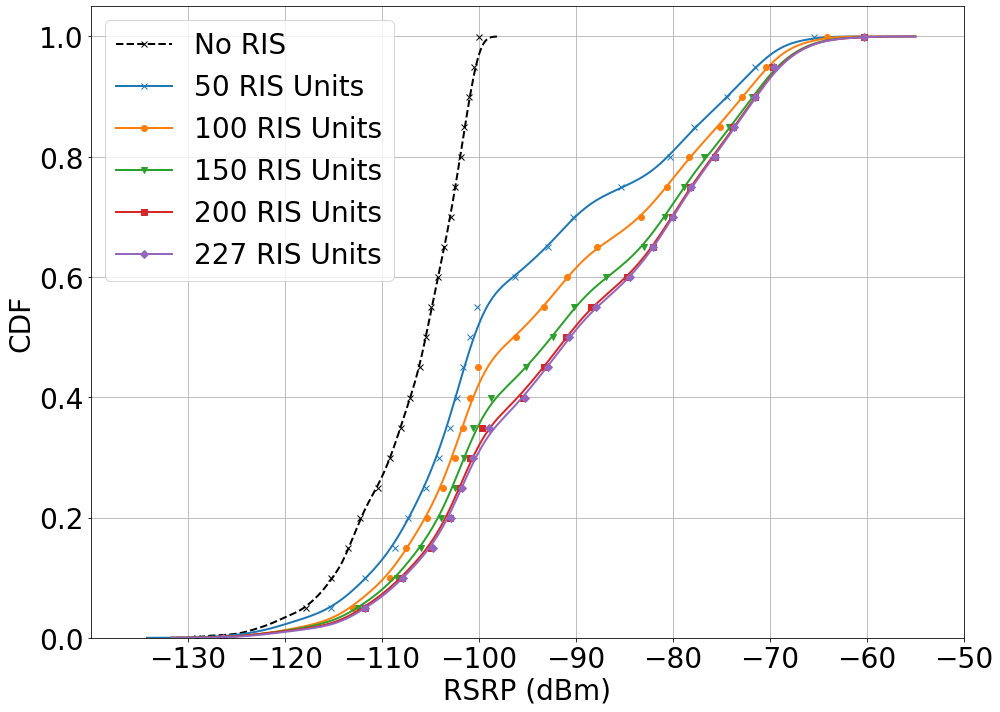}
        \label{fig:inc_RIS_3_5}
    }
    \hfill
    \subfloat[6G at 10\,GHz]{%
        \includegraphics[width=0.32\textwidth]{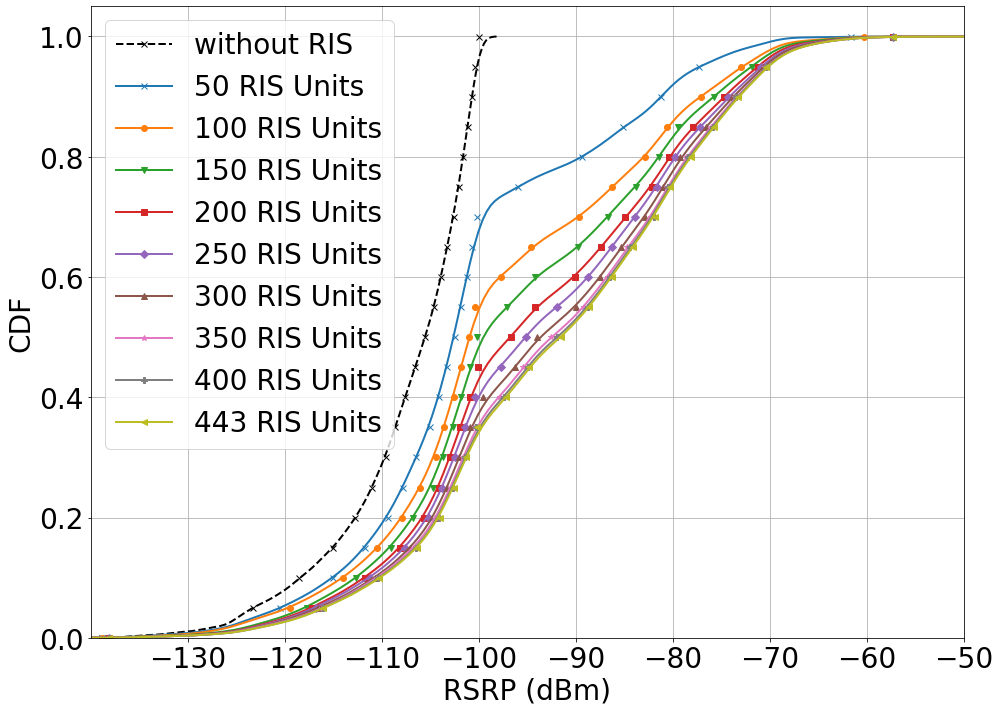}
        \label{fig:inc_RIS_10}
    }
    \caption{RSRP enhancement vs. number of deployed RIS.}
    \label{fig:Impact_RIS_incresing}
\end{figure*}

\subsubsection{4G at 2 GHz} Outage UEs account for 2.35\% of the total UE population. 
With \( T = 15 \), 293 clusters are formed, each being assigned one RIS. RIS placement brings into coverage 52.7\% of the UEs originally in outage. Re-clustering the remaining UEs with \( T = 10 \) m brings 5\% more into coverage, and RIS re-association contributes a further 6.9\%, for a total of 64.4\% of outage UEs recovered.
Shown in Fig.~\ref{fig:Optimization at 2GHz} is the CDF of the improved RSRPs. 

\subsubsection{5G at 3.5 GHz} Outage UEs comprise 1.9\% of the network. Using the same procedure, 46.8\% of outage UEs are addressed under RIS deployment. Re-clustering with \( T = 10 \) m adds 11.4\%, and RIS re-association yields an additional 6.8\%, bringing the total to 65\%. Figure~\ref{fig:Optimization at 3.5GHz} illustrates these improvements, with 227 RIS units deployed.

\subsubsection{6G at 10 GHz} At this frequency, 443 RIS units are deployed for the 5.15\% of UE originally in outage. RIS deployment brings 49.6\% of them into coverage, and re-clustering adds 2\%. RIS re-association is particularly effective here, adding 12.4\%, resulting in a total of 64.1\% of outage UEs recovered, as shown in Fig.~\ref{fig:Optimization at 10GHz}.

RIS gains behave differently across frequency bands, offering useful insights for practical deployment. At 2GHz, the improvements are modest and fairly uniform, suggesting that RIS mainly helps to smooth out weak coverage areas in a generally well-covered band. At 3.5GHz, the gains become stronger and more stable, showing that RIS can effectively handle moderate blockages while still benefiting from multipath propagation. At 10GHz, the improvements vary more widely—some UEs experience large gains while others benefit less—due to the higher sensitivity of high-frequency links to blockage. In this case, re-association has the most impact, emphasizing the importance of carefully matching UEs with RIS at higher frequencies.

These initial results indicate substantial coverage gains, but at the expense of an enormous RIS aperture and, with one RIS per cluster, a vast number of RIS units. The remainder of this section quantifies the impact of restricting the apertures and of limiting the number of RIS units to a subset of  clusters.

\subsection{Impact of RIS Density}
\label{subsec:impact_RIS_density}


In the foregoing evaluation, each cluster was assigned one RIS, for a total of 293, 227, and 443 RIS units at 2, 3.5, and 10\,GHz, respectively. However, many clusters are small: over 50\% contain fewer than six UEs (see Fig.~\ref{fig:user_bar_distribution}).
Next, clusters are sorted by UE count and RIS are incrementally assigned to the top-$N$ clusters, varying $N$. Figures~\ref{fig:inc_RIS_2},~\ref{fig:inc_RIS_3_5}, and~\ref{fig:inc_RIS_10} present the CDF of RSRP as a function of $N$,
confirming that prioritizing clusters by size does balance coverage improvements with infrastructure costs. Nonetheless, even with prioritization, the numbers of RIS units required to achieve substantial gains remain large. 

To evaluate whether the number of required RIS units could be reduced, we extended our framework to allow each RIS to also support nearby clusters within a 60\,m range. In this additional simulation, conducted for the 5G deployment at 3.5\,GHz with an RIS aperture of 11.5\,m~$\times$~11.5\,m, each RIS was configured to steer its beams toward all outage UEs within range, regardless of cluster membership. The deployed RISs were ranked by the number of UEs they supported and progressively activated—50, 100, 150, and so on—to assess the recovery of outage UEs as a function of the number of RIS units. As shown in Table~\ref{tab:ris_recovery}, allowing RISs to serve nearby clusters improved coverage slightly: with 50 deployed RISs, 48.2\% of outage UEs were recovered, and with 100 RISs, this increased to 61.4\%, about 3\% points higher than the cluster-limited case. However, these results indicate that serving nearby clusters offers limited benefit in dense urban environments, where buildings and obstacles already constrain coverage. These findings confirm that the number of deployed RIS units in our framework is realistic and not excessive for such environments.

\begin{table}[ht]
    \centering
    \caption{Recovered outage UEs vs. number of deployed RIS units}
    \label{tab:ris_recovery}
    \footnotesize
    \renewcommand{\arraystretch}{1.2}
    \setlength{\tabcolsep}{6pt}
    \resizebox{\columnwidth}{!}{%
    \begin{tabular}{|c|c|c|}
        \hline
        \textbf{\begin{tabular}[c]{@{}c@{}}Number of\\ deployed RIS units\end{tabular}} &
        \textbf{\begin{tabular}[c]{@{}c@{}}\% of recovered outage UEs\\ (RIS limited to its cluster)\end{tabular}} &
        \textbf{\begin{tabular}[c]{@{}c@{}}\% of recovered outage UEs\\ (RIS serving nearby clusters)\end{tabular}} \\
        \hline
        50  & 45.61~\% & 48.2~\% \\
        \hline
        100 & 58.8~\%  & 61.4~\% \\
        \hline
        150 & 62.9~\%  & 65.7~\% \\
        \hline
        200 & 64.3~\%  & 66.1~\% \\
        \hline
        227 & 65.4~\%  & 67.2~\% \\
        \hline
    \end{tabular}
    } 
\end{table}

Although the focus of this work is on coverage (RSRP), the results also have implications for energy efficiency. In particular, the FR3 deployment, characterized by wider inter-site distances than sub-6\,GHz bands, would normally require additional BSs to fill coverage gaps. The introduction of RIS enables these sparser deployments to maintain service continuity without activating new BS sites, thus reducing the network's overall power footprint and infrastructure cost. Nevertheless, achieving such savings still entails deploying a non-negligible number of RIS units.

\subsection{Impact of RIS Apertures}
\label{subsec:impact_RIS_aperture}

\begin{figure}[t]
    \centering
    \includegraphics[width=\columnwidth]{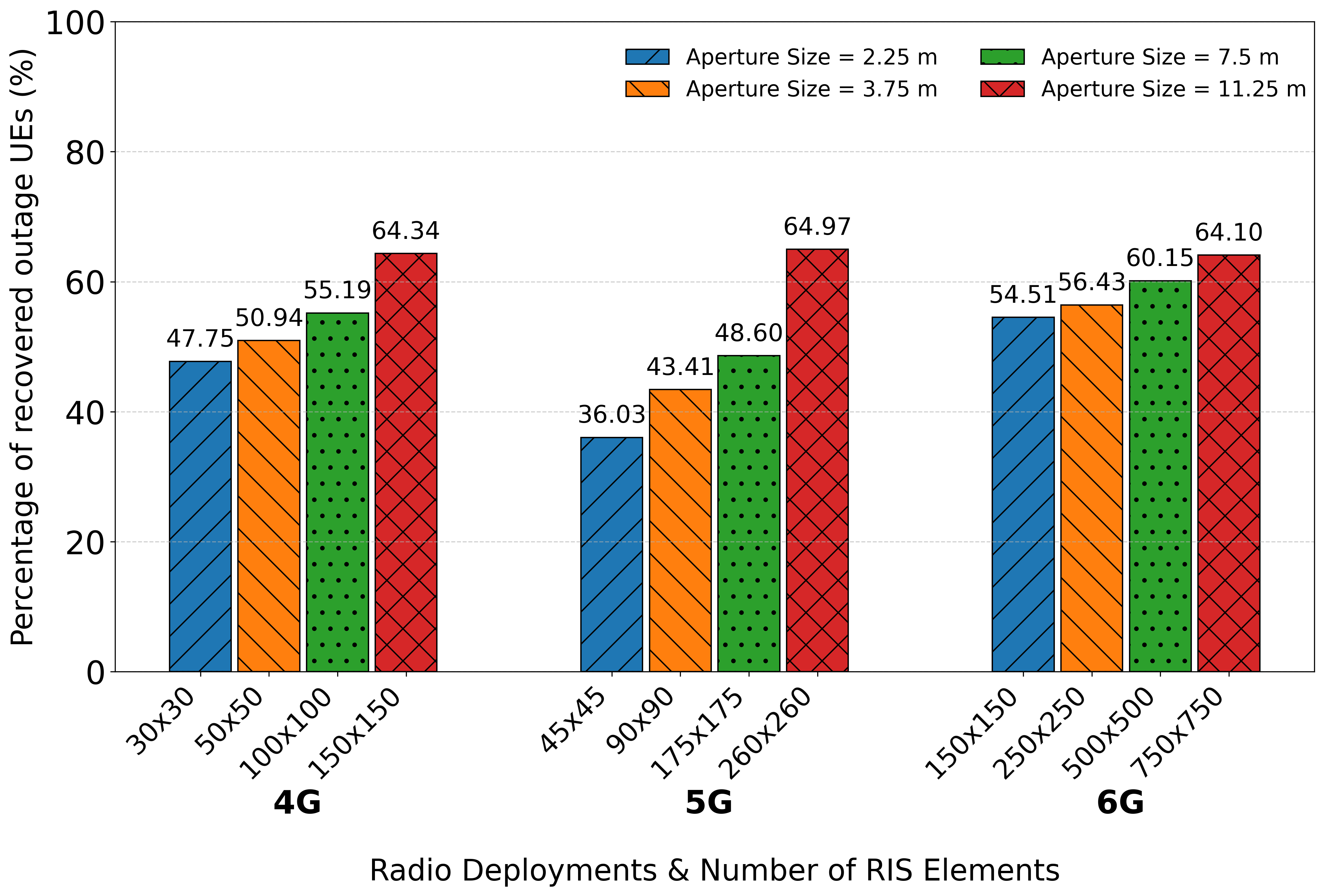}
    \caption{Percentage of outage UEs recovered with different RIS aperture sizes in various radio deployments.}
    \label{fig:ris_aperture}
\end{figure}

Next, let us investigate the effect of varying the aperture sizes, leading to different numbers of RIS elements. The element spacing is always half a wavelength. The assessment is carried out for the highest number of RIS units identified earlier, namely 293, 227, and 443 in 4G, 5G, and 6G, respectively. Figure~\ref{fig:ris_aperture} presents the percentage of outage UEs that are recovered, parameterized by the aperture size and with the corresponding number of RIS elements also indicated. The performance is seen to be more robust with respect to the aperture size than with respect to the RIS density, especially for 4G and 6G. 
 
\section{Validation with Calibrated Ray Tracing}
\label{sec:calibration}

Accurate ray tracing depends, besides precise geometric modeling, on a proper characterization of electromagnetic material properties.
This represents a major challenge, especially in large-scale environments. 
Previous studies have addressed this problem in controlled indoor settings or small outdoor scenes with a predefined ground truth~\cite{10705152, 10643616}. In contrast, this section focuses on material calibration in a dense urban environment, where default material libraries fail to reproduce the measured radio coverage.
A large dataset of outdoor wireless measurements collected in a large city in the UK is leveraged, corresponding to the same geographical area described in Sec.~\ref{sec:Sec2}. Specifically, our work relies on two complementary datasets: (i) a base station deployment dataset provided by the MNO, containing detailed configuration information such as antenna orientation, height, tilt, and beamwidth; and (ii) a crowdsourced measurement dataset collected via a proprietary mobile application disseminated to subscribers. Each measurement record comprises the RSRP and SINR values, the UE’s latitude and longitude at the time of reporting, and a flag indicating indoor or outdoor usage. Since our focus is on outdoor UEs, indoor samples are filtered out. In total, the dataset has approximately 8,000 samples within the calibrated area, ensuring sufficient statistical reliability for our analysis. These data allow calibrating material properties for the ray tracer. 

\begin{figure*}[t]
    \centering

    \subfloat[Scatter plot of simulated vs. measured RSRP]{%
        \includegraphics[width=0.42\textwidth]{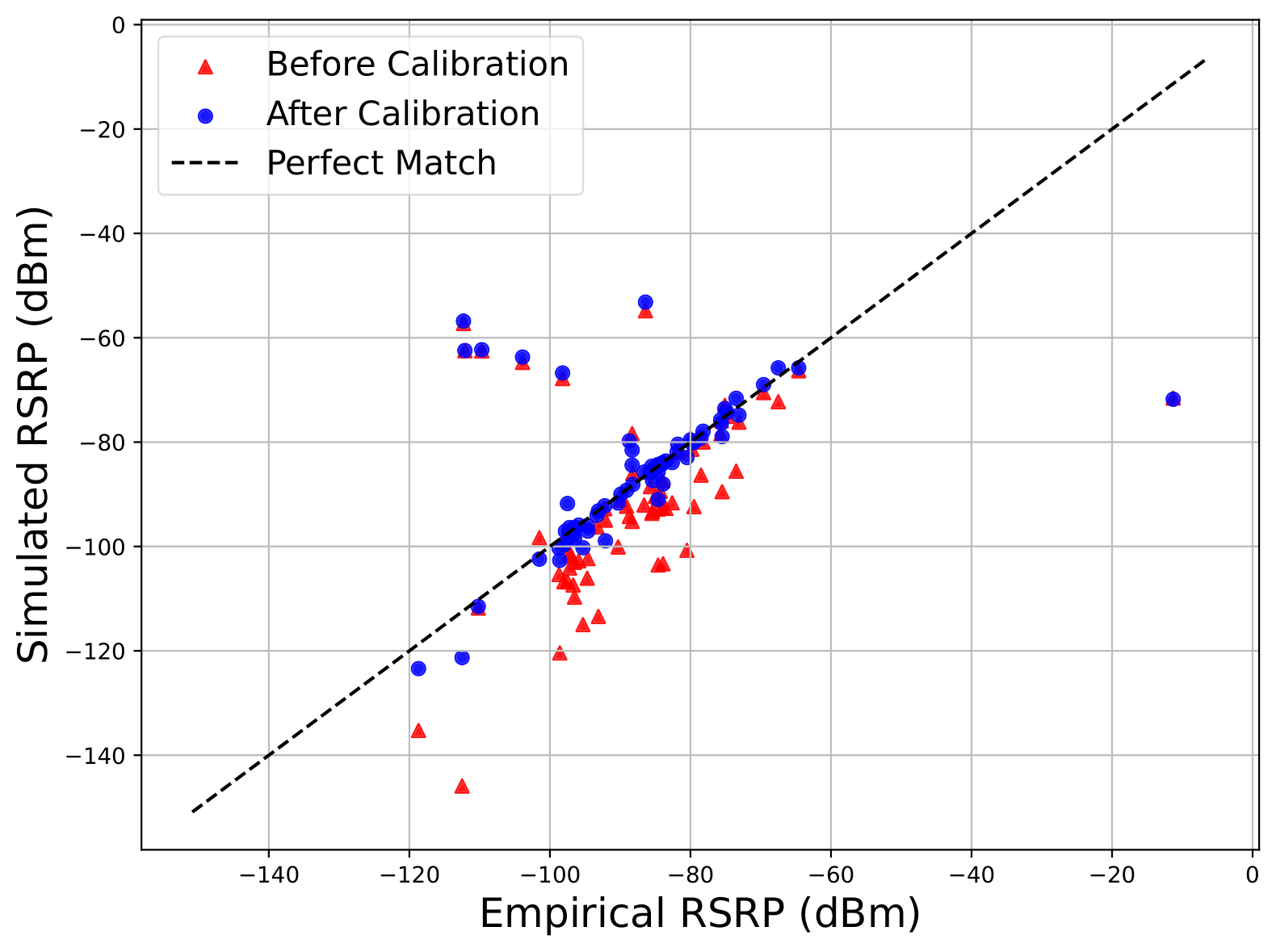}
        \label{cal_scatter}
    }
    \hfill
    \subfloat[Per-sample RSRP error]{%
        \includegraphics[width=0.5\textwidth]{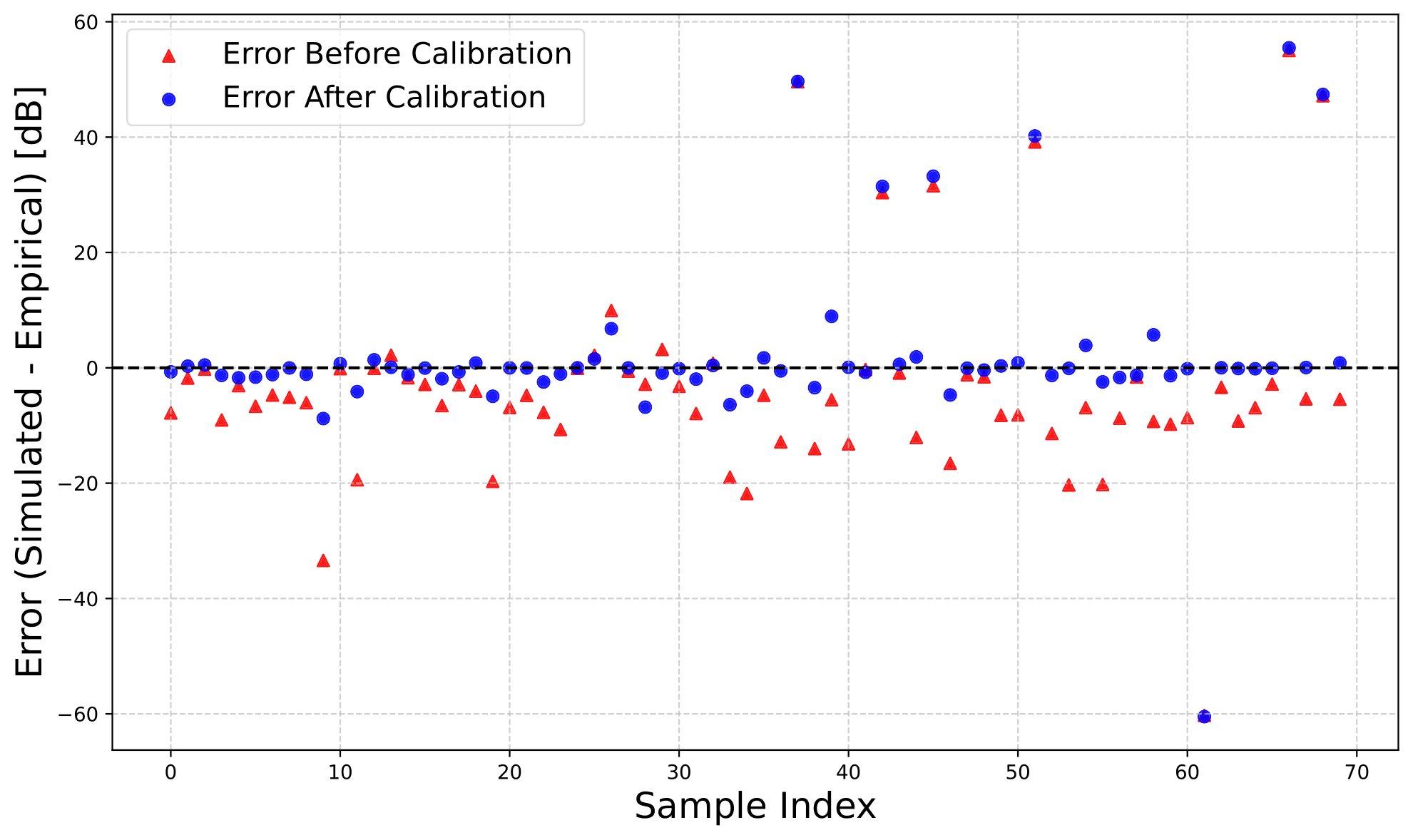}
        \label{cal_error}
    }

    \vspace{0.8em} 

    \subfloat[Simulated RSRP vs. empirical distribution]{%
        \includegraphics[width=0.42\textwidth]{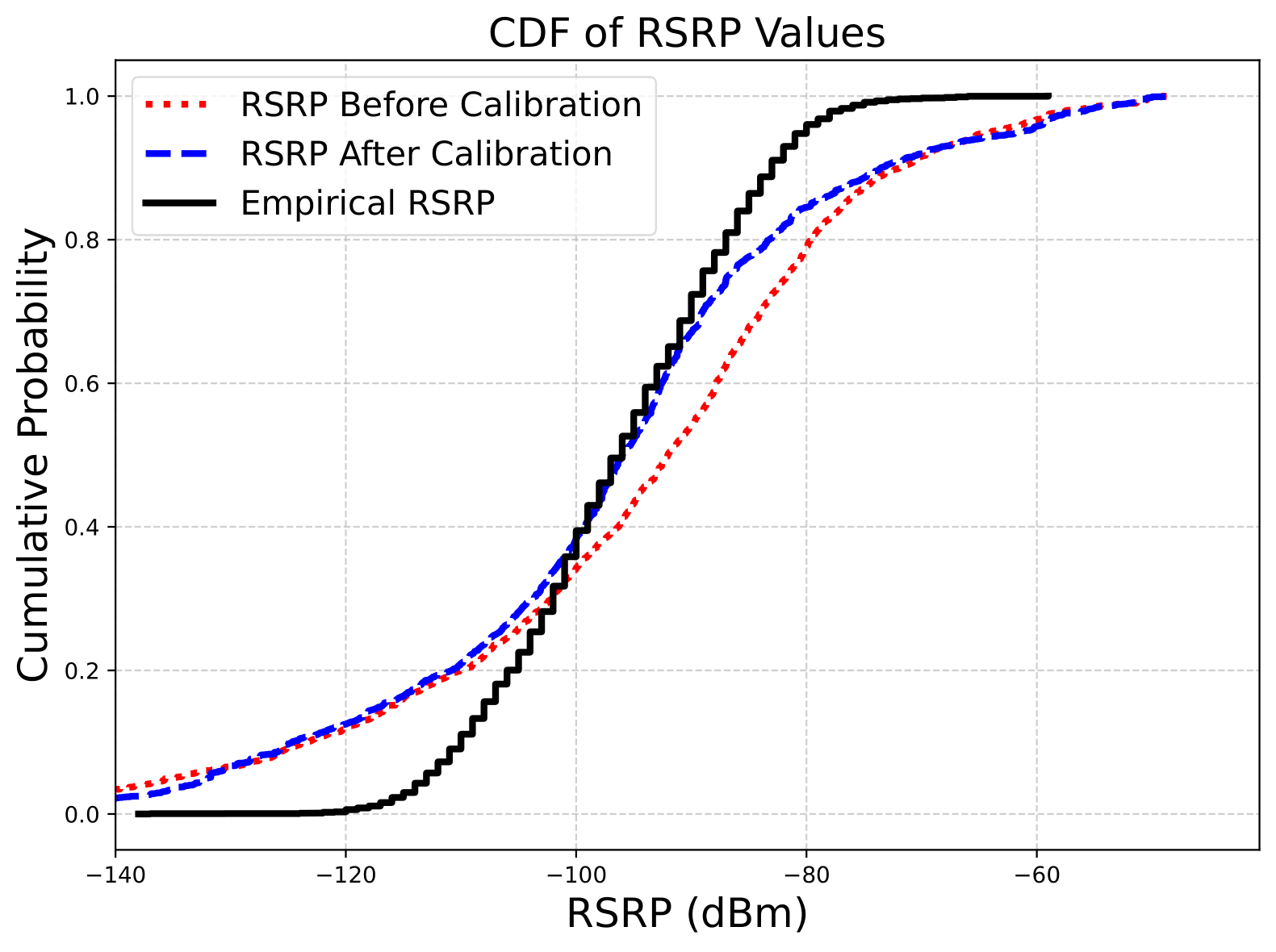}
        \label{cal_CDF}
    }
    \hfill
    \subfloat[Distribution of RSRP prediction errors]{%
        \includegraphics[width=0.5\textwidth]{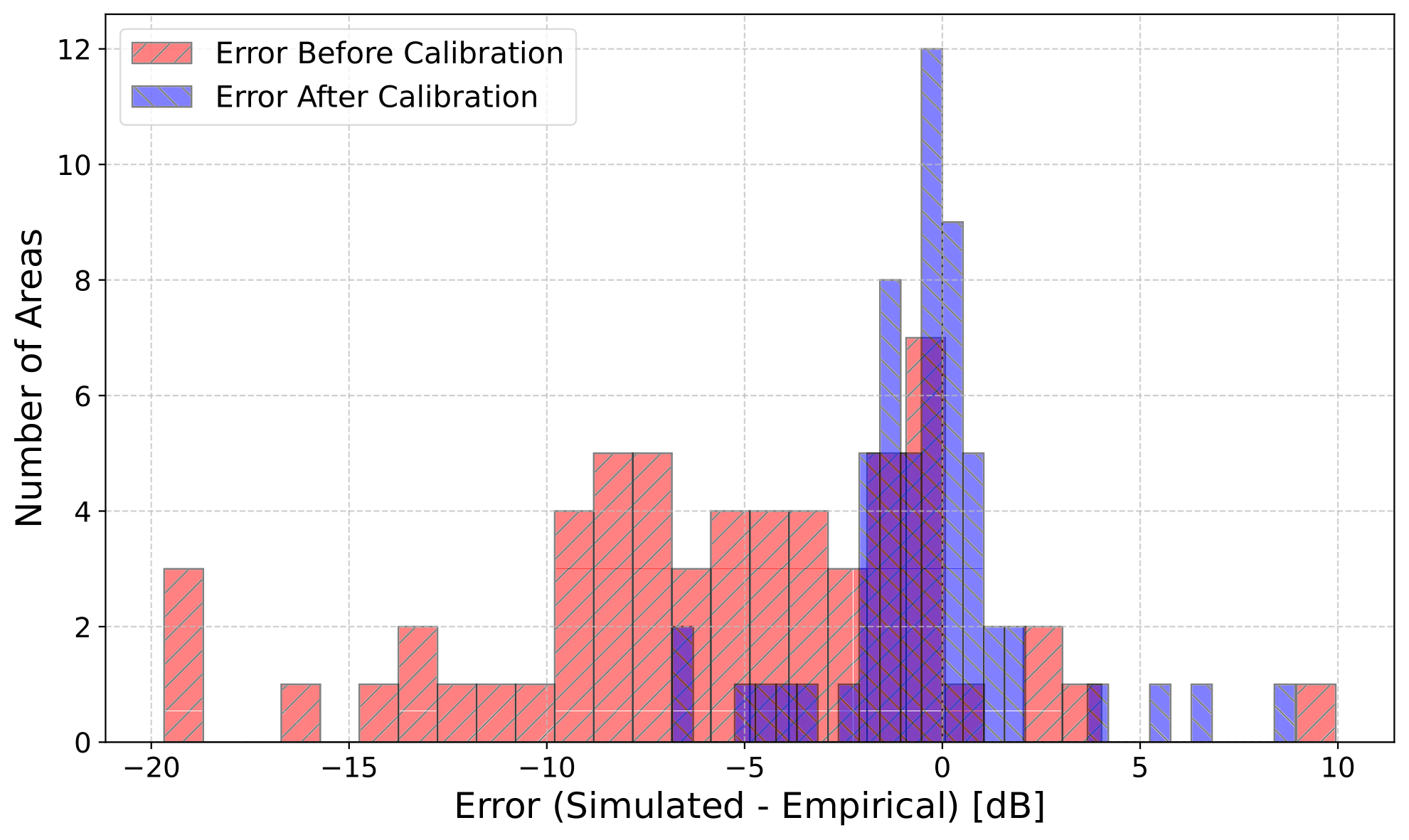}
        \label{cal_histogram}
    }

    \caption{Validation of the calibration, analyzing quantities before and after calibration: (a) Scatter plot of simulated vs. measured RSRP; (b) Per-sample RSRP error; (c) Simulated RSRP vs. empirical distribution; and (d) Distribution of RSRP prediction errors.}
    \label{fig:CDF_Comparison}
\end{figure*}

\subsection{Methodology}

The material properties are calibrated by minimizing the discrepancy between simulated and measured RSRP values. After importing the BS configuration into the Sionna engine, an initial coverage map is obtained using default material settings that assign concrete to all building surfaces. The following material parameters are treated as learnable variables during the optimization: relative permittivity (\( \epsilon \)), conductivity (\( \sigma \)), and a scattering coefficient (\( S \)) representing surface roughness.
The iterative, fully automated calibration procedure follows these steps:
\begin{itemize}
    \item UE measurement data is spatially mapped onto the simulation scene on a cell by cell basis. Target regions defined as areas of \(10 \times 10\,\mathrm{m}^2\) containing a minimum of 20 UE samples. These regions serve as the spatial units to ensure reliable statistical comparison between simulation and measurement. 
    
    \item The main scene is imported with the default materials for all buildings, which is concrete. All buildings within 100 meters of a target region (selected in the previous step) are automatically grouped and assigned a shared set of learnable material parameters based on the captured rays in the target regions. Buildings outside these regions retain default material properties (concrete). Initial parameter values are set to \( \epsilon = 5 \), \( \sigma = 5 \), and \( S = 0.5 \) and confined to the intervals \([1, 20]\), \([0, 15]\), and \([0, 1]\), respectively.
    
    \item At each iteration, a target region is randomly selected. 
    Its average simulated RSRP is compared to its average measured value, and the loss is defined as the error between the two.
 Gradients are computed with respect to the material parameters, and updates are performed via Adam optimizer with a learning rate of 0.05.
    
    \item This optimization runs for 600 iterations per cell, after which the next cell and its UEs are introduced. Calibrated building material are retained, and only uncalibrated buildings are treated as learnable in subsequent iterations, allowing localized calibration across overlapping regions.
\end{itemize}

The rationale behind this strategy stems from both modeling assumptions and data limitations:

\begin{itemize}
    \item As the goal is calibration, the absolute physical values of the parameters are secondary. The optimizer infers values consistent with the model that minimize the discrepancy between simulated and measured RSRP.
    
    \item The calibration takes place on the basis of target regions rather than individual measurements, to reduce the effects of UE variability. As UE measurements are collected over time and from different device manufacturers/models, per-UE RSRP can vary significantly.
    Averaging over at least 20 UEs ensures more stable values.
    
    \item Target regions with initial RSRP discrepancies exceeding 25\,dB between measurement and simulation are declared outliers---such mismatches are associated with missing geometry---and excluded.
    
    \item If the simulated RSRP significantly (see last bullet point)
exceeds the measured value, the optimizer pushes parameters toward free-space behavior (\( \epsilon \to 1 \), \( \sigma \to 0 \), \( S \to 0 \)). If the loss remains high at these extreme values, the region is excluded from further calibration.
    
    \item Conversely, if measured RSRP exceeds the simulated value (see last bullet point), the optimizer attempts to increase reflectivity by adjusting material parameters. If the discrepancy persists at the extreme values (\( \epsilon \to 20 \), \( \sigma \to 15 \), \( S \to 1 \)), the region is likewise excluded. These filtering steps ensure that calibration focuses on regions where meaningful parameter adjustments can be inferred.

    \item The considered urban environment features dense, tightly packed buildings. To manage simulation complexity, adjacent small structures are merged into unified building volumes, reducing polygon count and improving scalability of the ray tracing engine.

    \item The optimizer’s robustness to mismatch was seen not to be uniform across regions. For example, calibration in low-density urban areas often failed when the RSRP error exceeded 12 dB, whereas high-density urban environments could tolerate discrepancies of up to 20 dB. This highlights the importance of adopting context-aware calibration strategies and region-specific error thresholds.
    
\end{itemize}


\subsection{Validation}

Material calibration was performed on a central portion of the simulation area, spanning \(1122\,\mathrm{m} \times 710\,\mathrm{m}\).
Two validation procedures were conducted.

\subsubsection{Region Validation}
 First, 70 target regions are selected. For each, the average measured RSRP is contrasted with simulated values before and after calibration. Figure~\ref{cal_scatter} presents a scatter plot of simulated vs. measured RSRP, where a perfect match would result in points along the diagonal. As shown in Figure~\ref{cal_error}, the per-sample RSRP error decreased markedly post-calibration, with only a few remaining outliers due to geometry mismatches or measurement noise.

\subsubsection{RSRP Distribution}
Second, the CDF of RSRP was produced over all available UE measurements. Simulated RSRP before and after calibration is compared to the empirical distribution in Fig.~\ref{cal_CDF}. The pre-calibration CDF exhibits a clear deviation from the measured curve for all RSRP values, while the post-calibration result shows strong alignment in the range \(-105\) to \(-87\,\mathrm{dBm}\), where most UE data is concentrated. 
Remaining discrepancies in the lower and upper tails of the distribution can be explained by propagation characteristics. In the lower RSRP region, signals are dominated by weak multipath components that are relatively unaffected by material properties. In the upper region, UEs are typically in LoS, with reflections and diffractions having minimal influence. 

\subsubsection{Error Distribution}
To further quantify the improvement, the distribution of RSRP prediction errors is analyzed in Fig.~\ref{cal_histogram} with outlier regions excluded. Before calibration, the errors are widely spread and biased toward coverage underestimation, with a mean of \(-5.69\,\mathrm{dB}\), median of \(-5.37\,\mathrm{dB}\), and standard deviation of \(5.71\,\mathrm{dB}\). After calibration, the error distribution improves considerably, with a mean of \(-0.32\,\mathrm{dB}\), median of \(-0.13\,\mathrm{dB}\), and standard deviation of \(2.57\,\mathrm{dB}\).



\subsection{Performance of the Scattering-based Algorithm}

To wrap up this section, the scattering-based algorithm applied in the calibrated scene is compared with the reflection-based algorithm in the non-calibrated scene. To that end, the 4G radio deployment, along with its corresponding RIS and clustering specifications, is considered as outlined in Sec.~\ref{subsec:per_reflect}. Candidate RIS locations are identified based on single-bounce scattering rays reaching the centroid. The candidate location with minimum 3D distance to the centroid tile is selected. After optimizing the beamforming at the associated BS and the RIS phase configuration, the RSRP is evaluated at the centroid tile.
If the RSRP therein improves with the deployed RIS, the unit is configured and evaluated across the cluster UEs. A RIS is retained if more than 40\% of UEs benefit; otherwise, re-clustering is performed with a tighter threshold (\( T = 10 \)) to generate smaller, more localized clusters. The algorithm is reapplied to these new groups, and any remaining uncovered UEs are reassigned via the RIS re-association method described in Sec.~\ref{sec:reclustering}. 
The scattering-based approach recovers 70.2\% of the outage UEs after the initial clustering. Re-clustering increases this percentage to 73.2\%, and RIS re-association raises the total to 79.6\%. In this process, a total of 246 RIS units were deployed across the calibrated urban area. 
The results are qualitatively consistent with those obtained in the previous section using uncalibrated ray tracing.

\begin{figure}[t]
    \centering
    \includegraphics[width=\columnwidth]{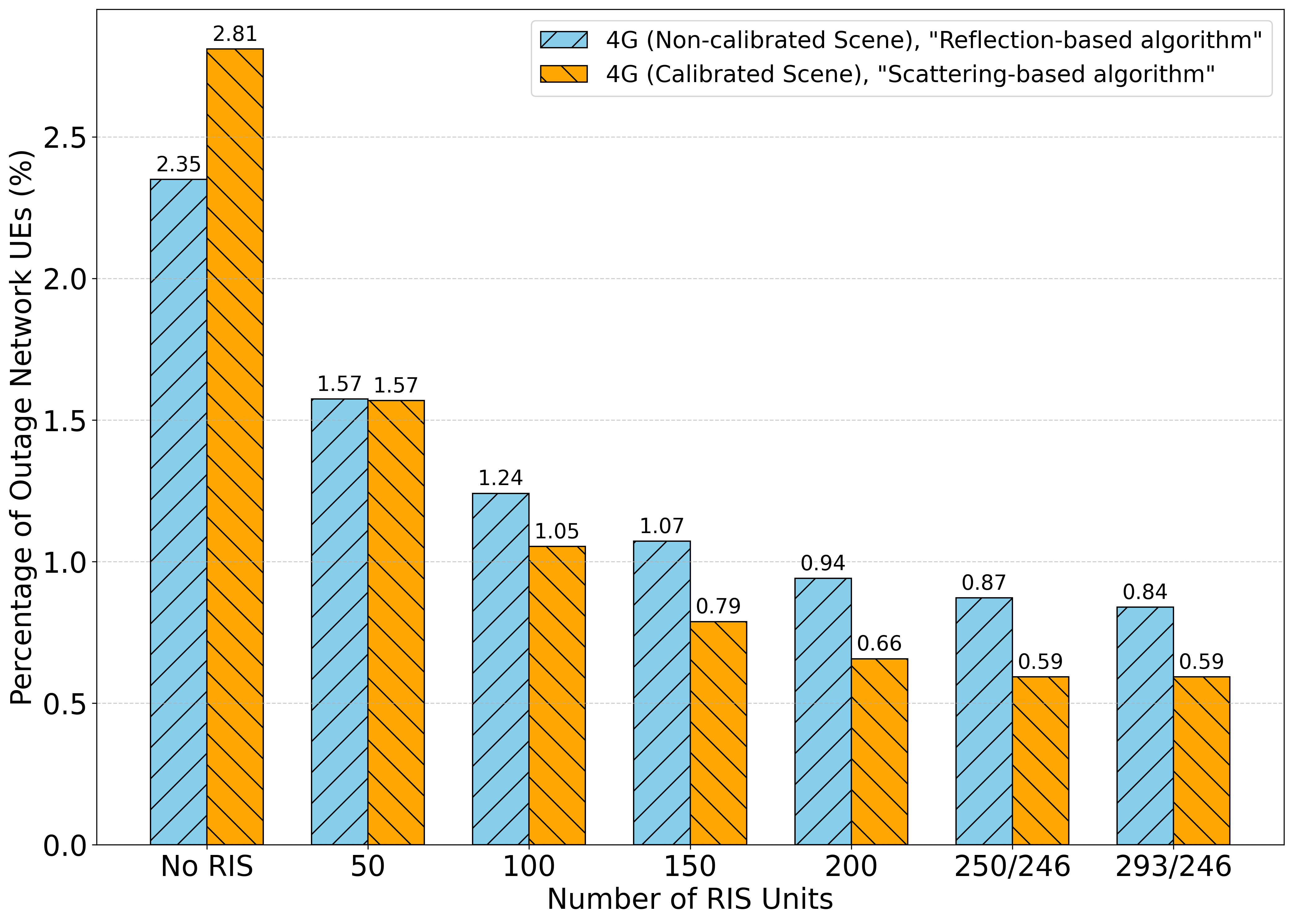}
    \caption{Percentage of outage UEs as a function of the number of deployed RIS units for a 4G deployment at 2\,GHz: non-calibrated ray tracing with reflection-based algorithm (blue) vs. calibrated ray tracing with scattering-based algorithm (orange).}
    \label{fig:alg_ris__incre_impro_Calib}
\end{figure}

Figure~\ref{fig:alg_ris__incre_impro_Calib} shows the percentage of outage UEs vs. the number of RIS units.
In the calibrated scene, the scattering-based algorithm achieves better performance with fewer RIS units (246) than the reflection-based algorithm in the non-calibrated scene (293). However, this still represents a major infrastructure investment. 

\section{Conclusion}
\label{sec:conclusion}

A fully automated, data-driven framework has been set forth to evaluate the effectiveness of large-scale RIS deployments in cellular networks. It integrates site-specific ray tracing, clustering of outage users, and ray-based heuristics to jointly determine RIS placement, orientation, phase configuration, and BS beamforming. Two complementary strategies are featured to identify effective RIS locations: based on specular reflections, or on scattering paths.
This framework has been evaluated in the context of 4G, 5G, and 6G systems modeled on a real-world urban deployment. To support the robustness of the findings, a material calibration procedure was further applied based on empirical signal measurements. This work provides open access to the full simulation framework, integrating ray tracing-driven RIS optimization algorithms, to foster reproducibility, benchmarking, and continued exploration. \footnote{Available at: \url{https://github.com/Telefonica-Scientific-Research/DDRD}} 

Shedding light on the tradeoff between performance and deployment costs, the presented evaluations indicate that achieving substantial improvements requires deploying dozens of large RIS units per square kilometer.
This is consistent with results in \cite{chizhik2022comparing,sadeghian2023ris} suggesting that, to have a substantial impact in the face of ambient reflections, scattering, and diffraction, RIS units would have to be exceedingly large.
It is also consistent with the findings in \cite{9999288}, where a trial in a commercial 5G network demonstrated only modest improvements with a single RIS.

Altogether, while a role may develop for the RIS in niche outdoor settings or perhaps indoors, this technology does not appear to be a cost-effective solution for broad outdoor coverage enhancements in dense urban areas within next-generation networks. Echoing this conclusion, 3GPP  has very recently discontinued its study item on the RIS for upcoming releases in favor of network-controlled repeaters \cite{aastrom2024ris}.
Being active, such repeaters can provide vastly superior power gains (up to 100 dB) without the need for proportionally large apertures, even if at the expense of the need for power supplies.

Notwithstanding the shortcomings of the RIS as a coverage enhancement mechanism, other potential benefits obtainable from the RIS do warrant continued research, say its ability to magnify spatial multiplexing in near-field multiantenna communication \cite{do2022line,do2022dof} or its potential for over-the-air computation and signal processing \cite{11095320}.



\bibliographystyle{IEEEtran}
\bibliography{journalAbbreviations, main}

\end{document}